\documentclass[aps,twocolumn,prl,floatfix,citeautoscript,nofootinbib,superscriptaddress]{revtex4}
\usepackage{eurosym}
\usepackage{amsfonts}
\usepackage{mathrsfs}
\usepackage{color}
\usepackage{xspace}
\usepackage{subfigure}
\usepackage{longtable}
\usepackage{xspace}
\usepackage{multirow}
\usepackage{tabularx}
\usepackage{amsmath}
\usepackage{amssymb}
\usepackage{graphicx}
\usepackage{dcolumn}
\usepackage{bm}
\usepackage[colorlinks,urlcolor=blue,citecolor=blue,linkcolor=blue]{hyperref}
\usepackage[normalem]{ulem}
\usepackage{mathrsfs}
\usepackage{epstopdf}
\usepackage[level]{datetime}
\usepackage{verbatimbox}
\usepackage{threeparttable}

\begin{document}

\title{Extracting the Quantum Geometric Tensor of an Optical Raman Lattice by Bloch State Tomography}
\date{\today}

\author{Chang-Rui Yi}
\thanks{These authors contribute equally to this work.}
\affiliation{Hefei National Research Center for Physical Sciences at the Microscale and School of Physical Sciences, University of Science and Technology of China, Hefei 230026, China}
\affiliation{Shanghai Research Center for Quantum Science and CAS Center for Excellence in Quantum Information and Quantum Physics, University of Science and Technology of China, Shanghai 201315, China}
\affiliation{Hefei National Laboratory, University of Science and Technology of China, Hefei 230088, China}

\author{Jinlong Yu}
\thanks{These authors contribute equally to this work.}
\affiliation{Center for Theoretical Physics, Hainan University, Haikou 570228, China}
\affiliation{School of Science, Hainan University, Haikou 570228, China}
\author{Huan Yuan}
\affiliation{Hefei National Research Center for Physical Sciences at the Microscale and School of Physical Sciences, University of Science and Technology of China, Hefei 230026, China}
\affiliation{Shanghai Research Center for Quantum Science and CAS Center for Excellence in Quantum Information and Quantum Physics, University of Science and Technology of China, Shanghai 201315, China}
\affiliation{Hefei National Laboratory, University of Science and Technology of China, Hefei 230088, China}
\author{Rui-Heng Jiao}
\affiliation{Hefei National Research Center for Physical Sciences at the Microscale and School of Physical Sciences, University of Science and Technology of China, Hefei 230026, China}
\affiliation{Shanghai Research Center for Quantum Science and CAS Center for Excellence in Quantum Information and Quantum Physics, University of Science and Technology of China, Shanghai 201315, China}
\affiliation{Hefei National Laboratory, University of Science and Technology of China, Hefei 230088, China}
\author{Yu-Meng Yang}
\author{Xiao Jiang}
\author{Jin-Yi Zhang}
\affiliation{Hefei National Research Center for Physical Sciences at the Microscale and School of Physical Sciences, University of Science and Technology of China, Hefei 230026, China}
\affiliation{Shanghai Research Center for Quantum Science and CAS Center for Excellence in Quantum Information and Quantum Physics, University of Science and Technology of China, Shanghai 201315, China}
\affiliation{Hefei National Laboratory, University of Science and Technology of China, Hefei 230088, China}

\author{Shuai Chen}
\affiliation{Hefei National Research Center for Physical Sciences at the Microscale and School of Physical Sciences, University of Science and Technology of China, Hefei 230026, China}
\affiliation{Shanghai Research Center for Quantum Science and CAS Center for Excellence in Quantum Information and Quantum Physics, University of Science and Technology of China, Shanghai 201315, China}
\affiliation{Hefei National Laboratory, University of Science and Technology of China, Hefei 230088, China}
\author{Jian-Wei Pan}
\affiliation{Hefei National Research Center for Physical Sciences at the Microscale and School of Physical Sciences, University of Science and Technology of China, Hefei 230026, China}
\affiliation{Shanghai Research Center for Quantum Science and CAS Center for Excellence in Quantum Information and Quantum Physics, University of Science and Technology of China, Shanghai 201315, China}
\affiliation{Hefei National Laboratory, University of Science and Technology of China, Hefei 230088, China}

\begin{abstract}
In Hilbert space, the geometry of the quantum state is identified by the quantum geometric tensor (QGT), whose imaginary part is the Berry curvature and real part is the quantum metric tensor.
Here, we experimentally realize a complete Bloch state tomography to directly measure eigenfunction of an optical Raman lattice for ultracold atoms.
Through the measured eigenfunction, the distribution of the complete QGT in the Brillouin zone is reconstructed, with which the topological invariants are extracted by the Berry curvature and the distances of quantum states in momentum space are measured by the quantum metric tensor.
Further, we experimentally test a predicted inequality between the Berry curvature and quantum metric tensor, which reveals a deep connection between topology and geometry.
\end{abstract}

\maketitle
In quantum physics, quantum states are defined in Hilbert spaces, in which Berry phases profoundly reflect the fundamental geometric structure of Hilbert spaces \cite{BerryPhase}.
In 1989, Berry further pointed out that the geometry of Hilbert space is fully determined by the quantum geometry tensor (QGT) \cite{book}.
The QGT has an antisymmetric imaginary part, i.e. the well-known Berry curvature, which unveils the geometric phase difference of a wave fuction between two parameter points in the parameter space.
Its associated topological phenomena have been extensively studied in many systems, such as the Ahoranov-Bohn effect \cite{ABphase}, the quantum Hall effect \cite{RevModPhys.58.519}, topological insulators and topological superconductors \cite{topo1,topo2}.
Moreover, the QGT has a symmetric real part termed as the quantum metric tensor \cite{metric1,metric2}, which describes the distance between two quantum states in the parameter space.
The quantum metric tensor can give the famous Euler characteristic number \cite{ECN} and is of great importance on obtaining the quantum Fisher information \cite{QFI1,QFI2} as well as depicting quantum phase transitions \cite{PhysRevLett.99.100603}.

The broadening of topological knowledge from condensed matter physics has rendered the possibility of quantum control towards measuring the QGT.
Recently, several experiments have been conducted for the measurements of the QGT of two-level systems in a superconducting qubit \cite{QGT1}, an NV center qubit \cite{QGT2,yu2022experimental} and exciton-polaritons \cite{QGT3}.
Nevertheless, in lattice systems, the complete quantum geometric tensor of Bloch bands is yet to be measured.
Specially, in ultracold atoms, optical lattices have been developed as a highly controllable and tunable quantum simulator to investigate 
topological band physics \cite{topoband}, which offers an ideal platform for the measurement of the QGT \cite{Asteria2019}.
In addition, there are several tomography experiments in cold atom systems,
based on the dynamics after a projection onto flat bands~\cite{PhysRevLett.113.045303,topoColdatom1,Tarnowski2019}, based on projection on different momenta by fast acceleration~\cite{wilsonLines}, based on the interference of paths surrounding~\cite{atominterferometer} or crossing band closing points~\cite{Charles2022}, and based on off-resonant coupling to higher bands~\cite{PhysRevLett.118.240403}.
Whereas these tomography protocols probe Bloch states in optical lattices in different aspects, none of them focus on studying the quantum geometric tensor.

Here, we experimentally measure the complete QGT of a topological band using ultracold atoms in an optical lattice.
We implement a complete Bloch state tomography using a well-controlled momentum-transferred Raman laser pulse in a two-dimensional (2D) optical Raman lattice~\cite{realization2DSOC,W.S_realization2DSOC}.
Based on the Bloch state tomography, the eigenfunction of two-band model in the Raman lattices is directly measured through the expectation values of three Pauli matrices.
Thus, the momentum-resolved Berry curvature and the quantum metric tensor, relying only on the eigenfunction, are simultaneously reconstructed, which probes the complete geometry of Bloch states in the Raman lattices.
Meanwhile, the Chern number \cite{topo1} 
(quantum volume \cite{PhysRevB.104.045103,GoldmanSciPostPhys}) is obtained from the integral of the Berry curvature (the quantum metric tensor) over the full Brillouin zone.
An inequality proved in Refs.~\cite{PhysRevB.104.045103,GoldmanSciPostPhys} between the quantum volume and the Chern number is also tested experimentally.

\begin{figure}
  \centering
  \includegraphics[width=1\linewidth]{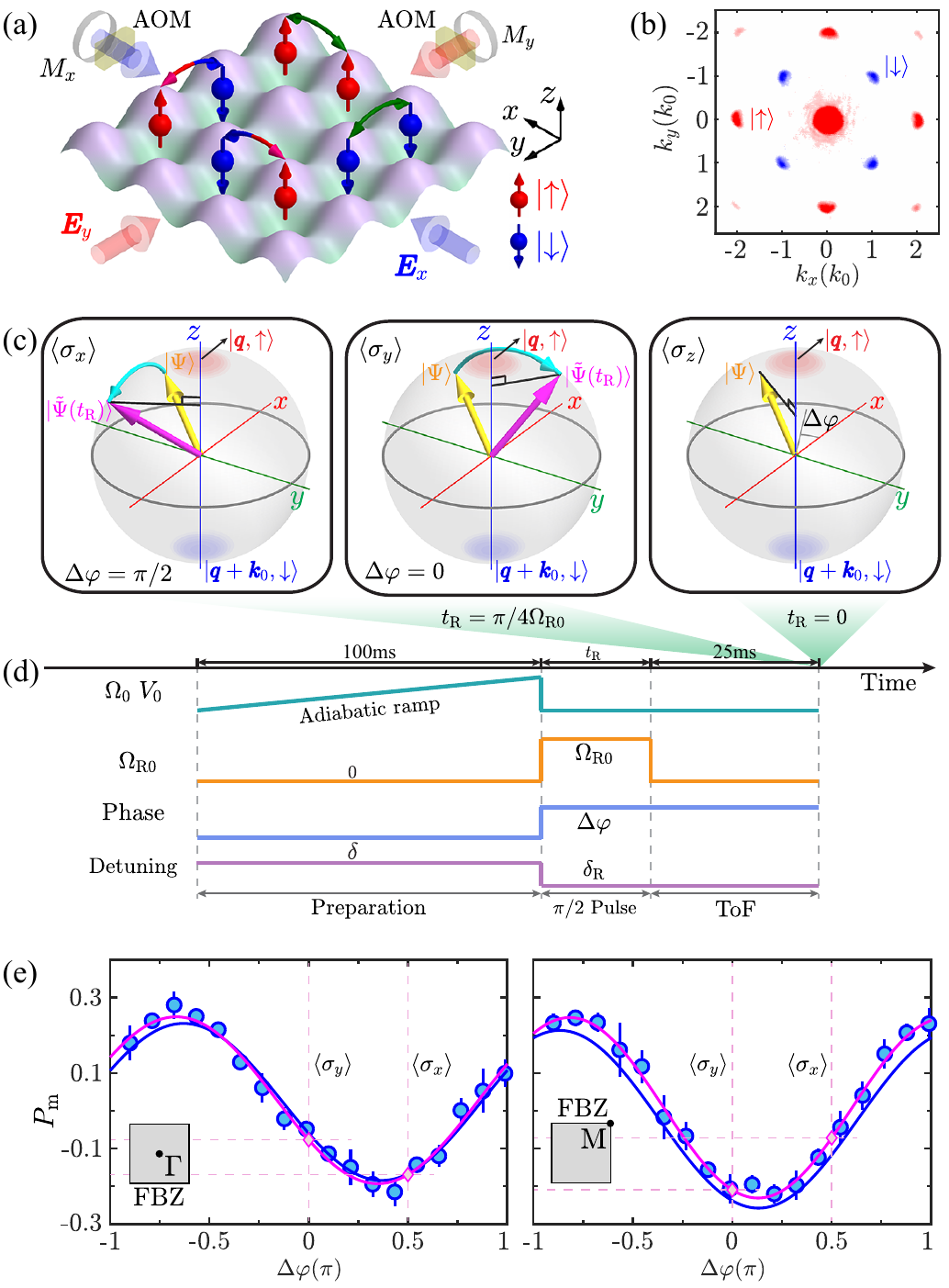}\\
  \caption{Bloch state tomography in 2D optical Raman lattices.
  (a) Sketch map of Raman lattices.
  (b) The distribution of Bose-Einstein condensates (BECs) in momentum space.
  BEC is prepared in $\left| \uparrow \right\rangle$ and stay at momentum point $\bm{k}=(0,0)$.
  Four BECs distributing at momenta $\bm{k}=(\pm k_0,\pm k_0)$ are in $\left| \downarrow \right\rangle$ due to
  the Raman potential.
  Here, $k_0$ is the recoil momentum.
  Other BECs are in $\left| \uparrow \right\rangle$ and diffracted by lattice potential without spin-flipping.
  (c) Tomography principle.
  The atoms firstly are in a superposition state $| \Psi \rangle$ of $\left| \bm{q},\uparrow \right\rangle$ and $\left| \bm{q}+\bm{k}_0,\downarrow \right\rangle$ (yellow arrows).
  Applying a $\pi/2$ Raman pulse, $| \Psi \rangle$ evolves to a new superposition state $| \tilde{\Psi} \rangle$ (magenta arrows) along the sky blue trajectories.
  $\Delta\varphi$ is exactly the angle between the gray axis and $x$-axis.
  The gray axis rotates in the equatorial plane by tuning $\Delta\varphi$.
  After ToF, $\left\langle \sigma_x \right\rangle$ ($\left\langle \sigma_y \right\rangle$) is obtained when $\Delta\varphi=\pi/2$ ($0$).
  $\left\langle \sigma_z \right\rangle$ is measured directly by ToF, without
  applying the Raman pulse.
  (d) The experimental sequence. The atoms are loaded in Raman lattices in the preparation stage.
  The Raman pulse is realized by suddenly turning off reflected beams of $\bm{E}_{x,y}$, setting Raman pulse strength to $\Omega_{\text{R0}}$, the relative phase $\Delta\varphi$ to a certain value and $\delta_{\text{R}}=0$.
  After ToF, the spin polarization is measured.
  (e) The spin polarization $P_{\text{m}}$ at the quasi-momentum $\Gamma$ and M as a function of the relative phase $\Delta \varphi$.
  The circles with error bar (blue curves) are from experimental data (numerical calculations).
  The experimental data is fitted by sinusoidal function (red curves).
  The insets mark $\Gamma$ and M points in the FBZ.
  Parameters: the lattice depth $V_0=4.0E_{\rm{r}}$, Raman potential strength $\Omega_0=1.0E_{\rm{r}}$, $\delta=-0.2E_{\rm{r}}$, $t_{\text{R}}=10\mu s$ and $\delta_{\text{R}}=0$.
  The recoil energy $E_{\rm{r}}\approx2\pi\times 3.7$kHz.
  }\label{Fig1}
\end{figure}

Our Bloch state tomography is realized in a 2D Raman lattice with ultracold ${}^{87}$Rb atoms~\cite{W.S_realization2DSOC,PhysRevLett.112.086401,supM}, as shown in Fig.\ref{Fig1}(a).
The Raman lattices are constructed by two laser beams $\bm{E}_{x,y}$ with wavelength $\lambda=787$nm that propagate in opposite directions as well as couple two magnetic levels $\left| F=1, m_{\rm F}=-1 \right\rangle \equiv \left|\uparrow \right\rangle$ and $\left|F=1, m_{\rm F}=0 \right\rangle \equiv \left|\downarrow \right\rangle$. The Hamiltonian is ($\hbar=1$)
\begin{equation}
H=\frac{\bm{k}^2}{2m}+V_{\rm latt}+\frac{\delta}{2}\sigma_z+\Omega_1\sigma_x+\Omega_2\sigma_y
,\label{Ham1}
\end{equation}
where $\bm{k}=(k_x,k_y), m, \delta$ and $\sigma_{x,y,z}$ are the momentum, mass of the atom,
two-photon detuning and Pauli matrices, respectively.
The Raman potential $\Omega_{1,2}$ (lattice potential $V_{\rm latt}$) induces spin-flipped (spin-conserved) hopping with the strength $\Omega_0$ ($V_0$) and transfers momentum of $\bm{k}_0=(\pm k_0,\pm k_0)$ ($2\bm{k}_0$), which is observed from the atomic distribution of ground state in the momentum space [Fig.\ref{Fig1}(b)]
\cite{realization2DSOC,W.S_realization2DSOC}.
Then, the eigenstate $\left| \Psi(\bm{q}) \right\rangle$ of Hamiltonian Eq.(\ref{Ham1}) can be denoted as the superposition of $\left|\bm{q},\uparrow \right\rangle$ and $\left|\bm{q}+\bm{k}_0,\downarrow \right\rangle$, i.e., $\left| \Psi(\bm{q}) \right\rangle=c_{\uparrow}(\bm{q})\left|\bm{q},\uparrow \right\rangle+c_{\downarrow}(\bm{q})\left|\bm{q}+\bm{k}_0,\downarrow \right\rangle$~\cite{supM}.
Here, $\bm{q}=(q_x,q_y)$ and $c_{\uparrow,\downarrow}$ are the quasi-momentum and the normalized complex coefficient, respectively.
The eigenstate $\left| \Psi(\bm{q}) \right\rangle$ can be represented on the Bloch sphere, whose poles are $\left|\bm{q},\uparrow \right\rangle$ and $\left|\bm{q}+\bm{k}_0,\downarrow \right\rangle$ [Fig.\ref{Fig1}(c)].

The main idea of the tomography is to measure the expectation values of three Pauli matrices $\left\langle \sigma_{x,y,z}(\bm{q}) \right\rangle$ for Raman lattices by rotating the measurement basis~(See~\cite{supM} for details).
We directly obtain the expectation value $\left\langle \sigma_{z}(\bm{q}) \right\rangle$ by spin-resolved time-of-flight (ToF) imaging, which has been applied in ~\cite{realization2DSOC,W.S_realization2DSOC}.
After rotating the measurement basis $| \bm{q},\uparrow \rangle$ and $| \bm{q}+\bm{k}_0,\downarrow \rangle$ to the $x$-axis ( $y$-axis) on the Bloch sphere, the expectation value $\left\langle \sigma_{x}(\bm{q}) \right\rangle$  ($\left\langle \sigma_{y}(\bm{q}) \right\rangle$) is observed by spin-resolved ToF imaging~\cite{PhysRevLett.107.235301}.
Such rotation is achieved by a coherent Raman pulse.
The pulse acting on $\left| \Psi(\bm{q}) \right\rangle$ requires transferring the momentum $\bm{k}_0=(-k_0,-k_0)$ between $\mid \bm{q},\uparrow \rangle$ and $\mid \bm{q}+\bm{k_0},\downarrow \rangle$ as well as maintaining a fixed relative phase $\Delta\varphi$ between the Raman pulse and the Raman lattices, which ensures accurate determination of $\left\langle \sigma_{x,y}(\bm{q}) \right\rangle$ [Fig.\ref{Fig1}(c)].

Concretely, the Raman pulse is made up of incident laser beams of $\bm{E}_{x,y}$ with retroreflective beams switched off by acousto-optic modulators  (AOMs) [Fig.\ref{Fig1}(a)].
Thus, the Hamiltonian of the Raman pulse is \cite{supM}
\begin{equation}
H_{\text{R}}=\begin{pmatrix}
  \bm{k}^2/2m+\delta_{\text{R}}/2& \Omega_{\text{R}}\\
  \Omega_{\text{R}}^*&\bm{k}^2/2m-\delta_{\text{R}}/2
\end{pmatrix}.
\end{equation}
In the experiment, a sufficient short Raman pulse with two-photon detuning $\delta_{\text{R}}=0$ is applied such that the kinetic energy is ignored and Raman coupling $\Omega_{\text{R}}$ dominates.
And $\Omega_{\text{R}}=\Omega_{\text{R0}}e^{-i[k_0(x+y)-\Delta\varphi]}$ with the strength $\Omega_{\text{R0}}$ launches a momentum transfer of $\bm{k_0}=(-k_0, -k_0)$ between $\mid \bm{q},\uparrow \rangle$ and $\mid \bm{q}+\bm{k_0}\downarrow \rangle$.
The relative phase $\Delta\varphi$ is controlled by the initial phase difference of the input lasers $\bm{E}_{x,y}$ that generates the Raman pulse and the Raman lattices \cite{supM}.

When the Raman pulse is applied to $\left| \Psi(\bm{q}) \right\rangle$ with the duration time $t_{\text{R}}$, the relative phase $\Delta\varphi$ is imprinted onto the time-dependent state $| \widetilde{\Psi}(\bm{q},t_{\text{R}}) \rangle$.
The state $| \widetilde{\Psi} \rangle$ precesses around an axis on the Bloch sphere with frequency $\Omega_{\text{R0}}$.
For instance, $| \widetilde{\Psi} \rangle$ precesses around $y$ ($x$)-axis in the equatorial plane when $\Delta\varphi=\pi/2 \;(0)$ [Fig.\ref{Fig1}(c)].
After ToF imaging,
the expectation values of Pauli matrices $\left\langle \sigma_{x,y,z} \right\rangle$ are obtained by the spin polarization~\cite{supM}
\begin{equation}\label{sigmat}
\begin{aligned}
\langle \widetilde{\Psi} | \sigma_z |\widetilde{\Psi} \rangle&=\left \langle \sigma_z \right \rangle \cos(2\Omega_{\text{R0}}t_{\text{R}})\\
&+(\left \langle \sigma_y \right \rangle \cos\Delta\varphi+
\left \langle \sigma_x \right \rangle\sin \Delta\varphi)\sin(2\Omega_{\text{R0}}t_{\text{R}}).
\end{aligned}
\end{equation}
When $t_{\text{R}}=0$, $\left\langle\sigma_{z}\right\rangle$ is obtained; when $t_{\text{R}}=\pi/(4\Omega_{R0})$, $\left\langle\sigma_{x}\right\rangle$ ($\left\langle\sigma_{y}\right\rangle$) is extracted by a $\pi/2$ Raman pulse with $\Delta\varphi=\pi/2$ ($0$) [Fig.\ref{Fig1}(c)].

\begin{figure}
  \centering
  \includegraphics[width=1\linewidth]{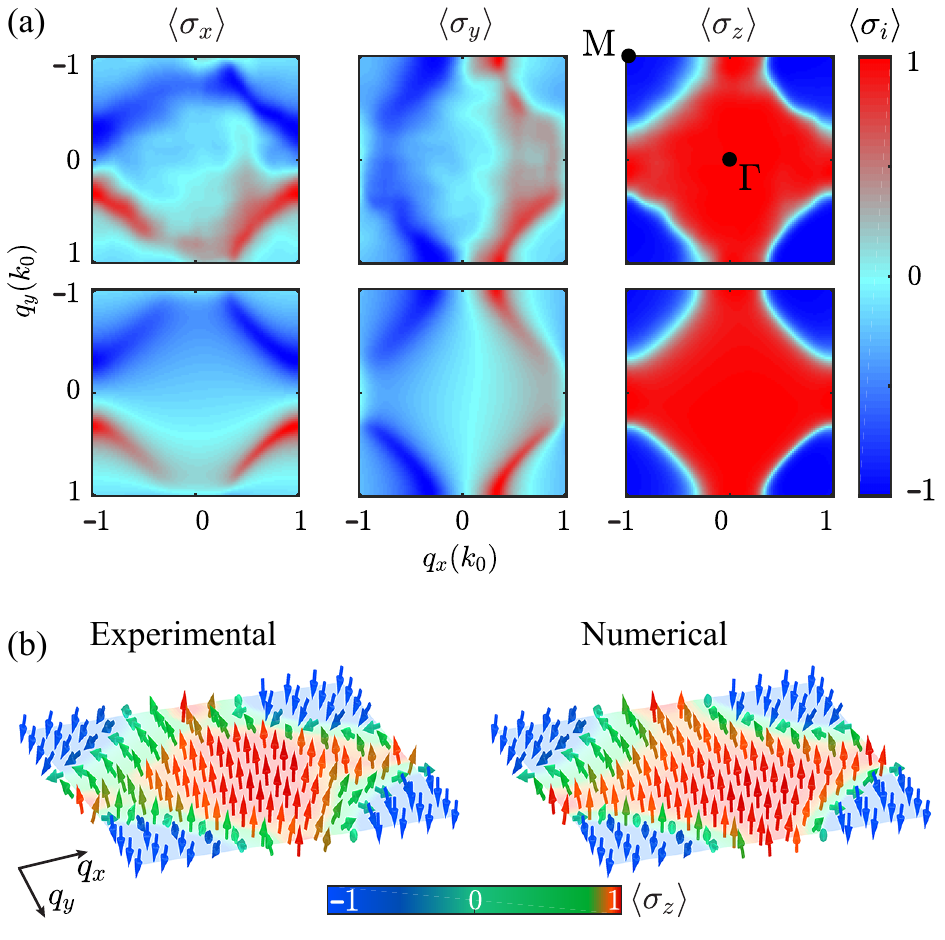}\\
  \caption{The expectation value of three Pauli matrices $\langle\sigma_{i=x,y,z}\rangle$ in the FBZ.
  (a) Normalized $\langle \sigma_{x,y,z} (\mathbf{q}) \rangle$ with $\delta=-0.2E_{\rm{r}}$ for the lowest band.
  Results from experimental data (upper row) are compared with numerical calculations (lower row).
  (b) The distribution of $\langle \bm{\sigma}(\mathbf{q}) \rangle$ in the FBZ takes a skyrmion configuration.
  Parameters: $(V_0,\Omega_0,\delta)=(4.0,1.0,-0.2)E_{\rm{r}}$.
  }\label{Fig2}
\end{figure}

Now, we clarify this technique in the experiment \cite{supM}.
The experimental protocol is depicted in Fig.\ref{Fig1}(d).
First, BEC of ${}^{87}$Rb atoms is adiabatically loaded into the Raman lattices.
The BECs condense at $\Gamma$ or M in the first Brillouin zone (FBZ), which depends on BEC prepared in $\left| \uparrow \right\rangle$ or $\left| \downarrow \right\rangle$.
Meanwhile, atoms stay at the lowest band of Raman lattices.
Second, the $\pi$/2 Raman pulse is switched on within 200ns to couple the atoms (See~\cite{supM} for the sensitivities of the pulse), which is accomplished by simultaneously executing the following manipulations on the beams $\bm{E}_{x,y}$: (i) Turning off the retroreflective beams of $\bm{E}_{x,y}$;
(ii) Turning on the strength of Raman pulse to $\Omega_{\text{R0}}\approx3.4E_{\rm{r}}$ by tuning the intensity of $\bm{E}_{x,y}$;
(iii) Setting the relative phase $\Delta\varphi$ to a certain value by changing the initial phase of $\bm{E}_{x,y}$;
(iv) Setting $\delta_{\text{R}}=0$ by adjusting the frequency of $\bm{E}_{x,y}$.
Finally, after holding the Raman pulse for a certain duration $t_{\text{R}}=10\mu s$, we measure the atomic numbers $n_{\uparrow,\downarrow}$ using spin-resolved ToF imaging to obtain spin polarization, i.e., $P_{\rm{m}}=(n_{\uparrow}-n_{\downarrow})/(n_{\uparrow}+n_{\downarrow})$.

The measured spin polarizations $P_{\rm{m}}$ versus the relative phase $\Delta\varphi$ are shown in Fig.\ref{Fig1}(e).
The spin polarizations at $\Gamma$ and M are fitted by 
sinusoidal functions, which demonstrates the superposition between $\left\langle \sigma_x \right\rangle$ and $\left\langle \sigma_y \right\rangle$ [cf., Eq.~(\ref{sigmat})].
Thus, $\left\langle \sigma_x \right\rangle$ ($\left\langle \sigma_y \right\rangle$) are extracted from the fittings when $\Delta\varphi=\pi/2$ ($0$) at $\Gamma$ and M, marked by diamonds.
For comparison, the numerical simulations of the spin polarizations with the same parameters are also plotted in Fig.\ref{Fig1}(e), in consistent with the experimental data.
All these indicate that Bloch state tomography in the Raman lattices is achieved.

Next, we explore geometry and topology with this novel Bloch state tomography to directly map out the quantum geometric tensor from the eigenfunction based on measuring $\left\langle \sigma_{x,y,z}(\bm{q}) \right\rangle$ of Bose gases in the FBZ.
In that case, almost identical procedure used in the aforementioned BEC measurement is employed, except that the atomic cloud is cooled to a temperature around $100$nK~\cite{supM}.
Then, the entire lowest band is occupied by a sufficient number of atoms; simultaneously,
around one third of the atoms populate to the higher bands,
which reduces the contrast of the spin polarizations being the same as Refs. \cite{realization2DSOC,W.S_realization2DSOC,realization3DSOC}.
Whereupon, we subtract the high-band contribution based on Bose distribution determined by the numerical calculations to obtain $\left\langle \sigma_{x,y,z}(\bm{q}) \right\rangle$ for the lowest band (See \cite{supM} for details).

\begin{figure}
  \includegraphics[width=1\linewidth]{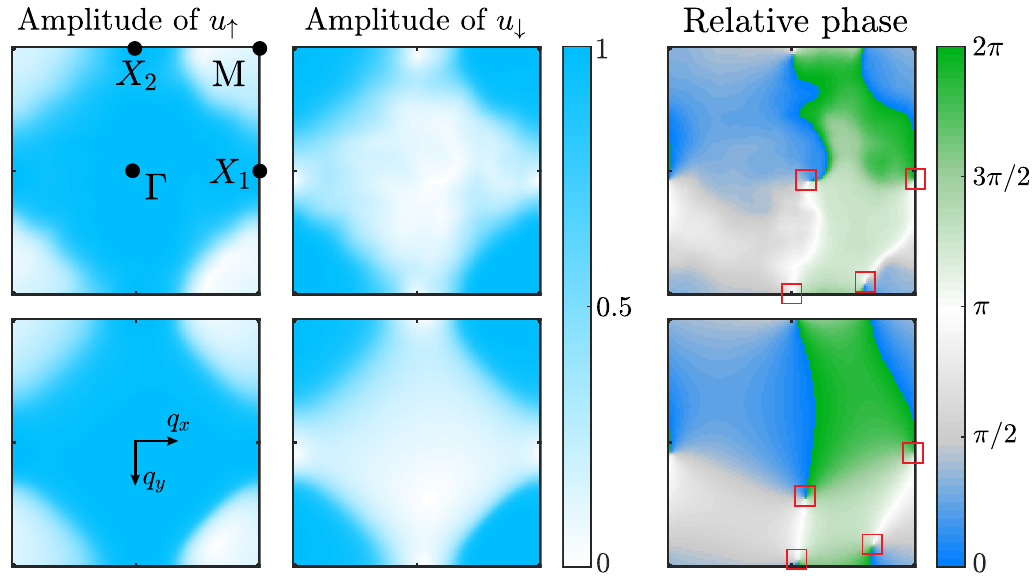}
  \caption{Experimental reconstruction of the eigenfunction in the FBZ: The amplitude of the eigenfunction $u_\uparrow$ (first column) and $u_\downarrow$ (second column), together with the relative phase between $u_\uparrow$ and $u_\downarrow$ (third column). Red squares mark the positions of the phase vortices.
  Results from experimental data (upper row) are compared with numerical calculations (lower row).
  Parameters: $(V_0,~\Omega_0,~\delta)=(4.0,1.0,-0.2)E_{\rm{r}}$.
  }
  \label{Fig3}
\end{figure}

The typical normalized $\left\langle \sigma_{x,y,z}(\bm{q}) \right\rangle$ are drawn in Fig.\ref{Fig2}(a), which can be divided into two regions with $\langle \sigma_{x,y,z} \rangle<0$ and $\langle \sigma_{x,y,z} \rangle>0$.
For $\langle \sigma_{x}(\bm{q}) \rangle$ ($\langle \sigma_{y}(\bm{q}) \rangle$), the two regions are distributed in the upper and lower (left and right) of the FBZ, respectively.
For $\langle \sigma_{z}(\bm{q}) \rangle$, one of the regions is centered on $\Gamma$ and the other region is centered on M for $\delta=-0.2E_{\rm{r}}$.
And the two regions are demarcated by a ring structure with $\langle \sigma_{z} \rangle=0$ around M, which is a feature of band topology in our system \cite{realization2DSOC,W.S_realization2DSOC}.
The numerical calculations coincide with the experimental measurements.
In addition, the distribution of the vectors $\langle \bm{\sigma}(\mathbf{q}) \rangle = (\left\langle \sigma_{x}(\bm{q}) \right\rangle, \left\langle \sigma_{y}(\bm{q}) \right\rangle, \left\langle \sigma_{z}(\bm{q}) \right\rangle)$ shapes a skyrmion \cite{Skyrme1962} structure in the momentum space, and the twisted skyrmion texture is demonstrated in Fig.\ref{Fig2}(b).
Such skyrmion configuration offers the direct evidence for complete tomography.
\begin{figure}
  \includegraphics[width=1\linewidth]{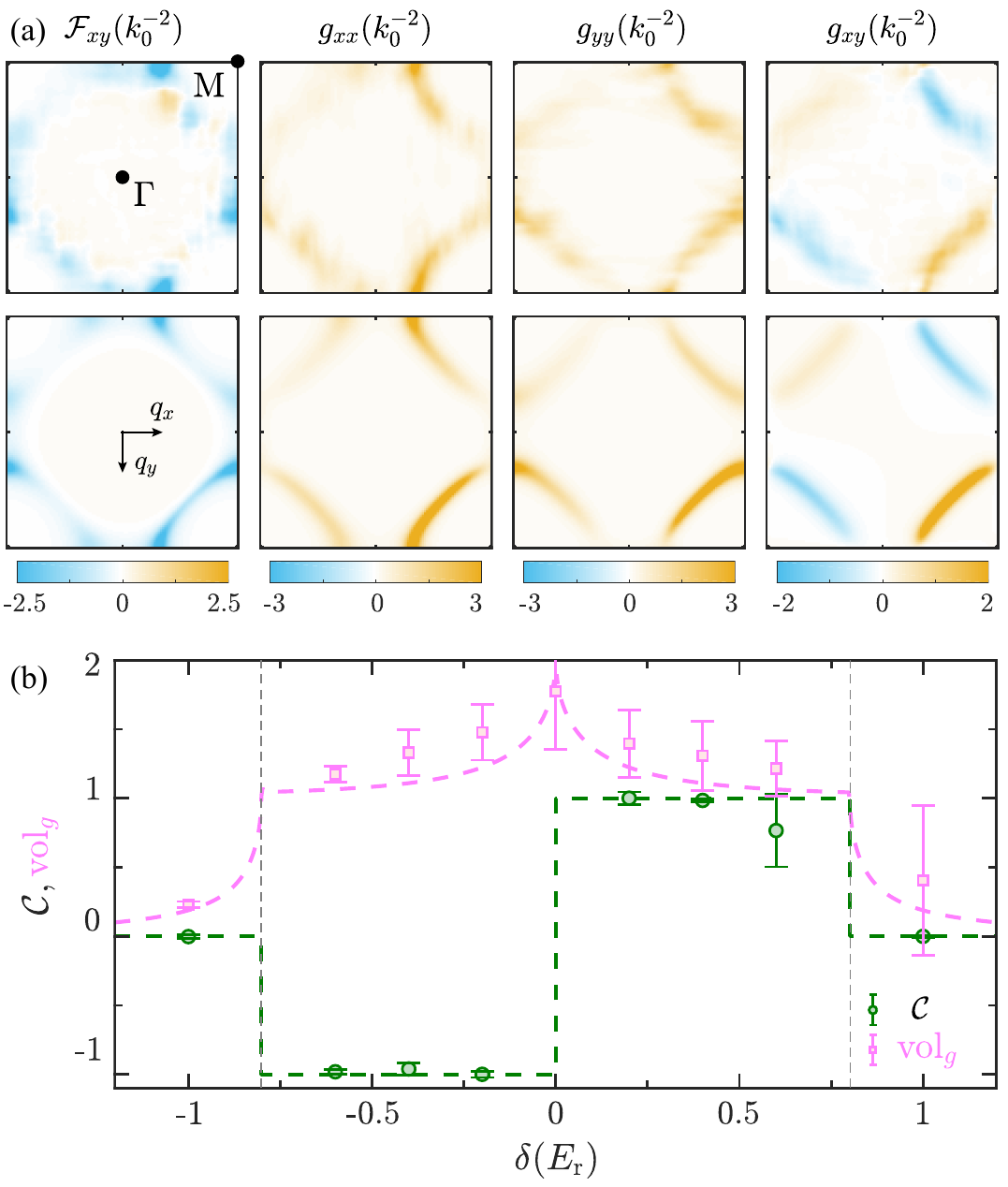}
  \caption{The quantum geometric tensor in the FBZ.
  (a) The Berry curvature $\mathcal{F}_{xy}(\bm{q})$, together with the three components of the quantum metric tensor $g_{xx}(\bm{q})$, $g_{yy}(\bm{q})$, $g_{xy}(\bm{q})$.
  Results from experimental data (upper row) are compared with numerical calculations (lower row).
  Parameters: $(V_0,~\Omega_0,~\delta)=(4.0,1.0,-0.2)E_{\rm{r}}$.
  (b) The Chern number $\mathcal{C}$ and the quantum volume $\text{vol}_g$ versus the detuning $\delta$.
  The circles and squares with error bars are from experimental results.
  The dash pink and green curves are from numerical calculations.
  Parameters: $(V_0,~\Omega_0)=(4.0,1.0)E_{\rm{r}}$.
  }
  \label{Fig4}
\end{figure}

Moreover, the eigenfunction can also be extracted from the normalized $\left\langle \sigma_{x,y,z} \right\rangle$.
To this end, we consider the Bloch Hamiltonian under the two-band tight-binding approximation of the Eq.(\ref{Ham1}) \cite{realization2DSOC,uncoverTopology}.
After diagonalizing the Bloch Hamiltonian, the eigenfunction in the lowest band can be written as $\left| u(\bm{q}) \right\rangle=\left( \sin(\theta_{\bm{q}}/2)e^{-i\phi_{\bm{q}}}, -\cos(\theta_{\bm{q}}/2) \right)^{\rm{T}}$, where the parameters $\theta_{\bm{q}}=\arccos(-\left\langle \sigma_z \right\rangle)\in[0,\pi]$ and $\phi_{\bm{q}}=\arg(\left\langle \sigma_x \right\rangle+i\left\langle \sigma_y \right\rangle)\in[0,2\pi)$.
The amplitudes of the eigenfunction $u_{\uparrow,\downarrow}$ and their relative phase for $\delta=-0.2E_{\rm{r}}$ in the FBZ are exhibited in Fig.\ref{Fig3}.
The amplitude of the eigenfunction $u_{\uparrow}$ has a value of $\sim$1 at the $\Gamma$ point and is close to zero at the M point, whereas the amplitude of $u_{\downarrow}$ is opposite.
The relative phase is significantly different from the amplitude and can be roughly divided into four regions: the blue-gray region ($0\leq \varphi < 0.5\pi$), the gray region ($0.5\pi\leq \varphi < \pi$) on the left of the phase map, and the light-green region ($\pi\leq \varphi < 1.5\pi$), the green region ($1.5\pi\leq \varphi < 2\pi$) on the right of the phase map.
The junctions of these four regions emerge phase vortices, which implies that the vectors $\langle \bm{\sigma} \rangle$ point to north or south poles of the Bloch sphere \cite{2017Yu}.
Our experimental measurements basically agree with numerical calculations (second row of Fig.\ref{Fig3}).
Note that the phase vortices deviate from the high-symmetry points of the FBZ \{$\Gamma$,M,$X_{1,2}$\}, which mainly stems from the coherence between the bands~\cite{supM}.

Thanks to the complete observation of the eigenfunction $|u(\bm{q})\rangle$, we can identify the geometric structures of the quantum states by the gauge-invariant quantum geometric tensor $\chi_{\alpha\beta}$ \cite{QGTs}
\begin{equation}\label{QGT}
\chi_{\alpha\beta}=\langle \partial_{q_{\alpha}} u| (1-|u\rangle\langle u|)|\partial_{q_{\beta}}u\rangle=g_{\alpha\beta}-i\mathcal{F}_{\alpha\beta}/2.
\end{equation}
Here, the quantum metric tensor $g_{\alpha\beta}=\rm{Re}(\chi_{\alpha\beta})$, as the real part of QGT, measures a distance between quantum states in quasi-momentum space \cite{metric1}.
The Berry curvature $\mathcal{F}_{\alpha\beta}=-2\rm{Im}(\chi_{\alpha\beta})$, being the imaginary part of QGT, acts as an effective `electromagnetic' tensor in quasi-momentum space \cite{topo1}.
Hence, the momentum-resolved quantum metric tensor $g_{\alpha\beta}$ and Berry curvature $\mathcal{F}_{xy}$ are extracted from the measurement of eigenfunction in Fig.\ref{Fig3} using Eq.(\ref{QGT}), displayed in Fig.\ref{Fig4}(a).
The Berry curvature $\mathcal{F}_{xy}$ is mainly localized near the ring structure, the sign and direction of which is negative and perpendicular to the plane $q_x$-$q_y$, respectively.
Thereby, Chern number $\mathcal{C}=\int_{\text{BZ}}\mathcal{F}_{xy}d\bm{q}/2\pi=-1.00\pm0.02$ at $\delta=-0.2E_{\text{r}}$, signifying the band is topologically non-trivial.
Tuning the detuning from $\delta=-1.0E_{\text{r}}$ to $\delta=1.0E_{\text{r}}$, in Fig.\ref{Fig4}(b), the topological phase diagram is obtained by the Chern number, in consistent with our previous measurements via quench dynamics~\cite{uncoverTopology,topoCharge,windingNumber_zhang}.

In addition, the quantum metric tensor with non-vanishing value in Fig.\ref{Fig4}(a) also appears on the ring structure around M, indicating the distance between quantum states on the ring is larger than other areas.
Moreover, the integral of the quantum metric tensor over the full Brillouin zone gives a quantum volume~\cite{note1} $\text{vol}_g=\int_{\text{BZ}}\sqrt{g_{xx}g_{yy}-g_{xy}^2}d\bm{q}/\pi$, which yields an inequality $\text{vol}_g\ge |\mathcal{C}|$ for Chern insulators~\cite{PhysRevB.104.045103,GoldmanSciPostPhys}.
Figure \ref{Fig4}(b) shows the quantum volume and the Chern number as functions of the detuning, validating such an inequality experimentally.
The inequality is related to superfluidity or superconductivity~\cite{Peotta2015}.
Thereafter, we may use the inequality to roughly evaluate the topological properties in 2D Raman lattices: If $\text{vol}_g<1$, $\mathcal{C}=0$; If $\text{vol}_g\ge 1$ and $\delta\neq 0$, $|\mathcal{C}|=1$.
Note that the quantum volume takes non-integer value that can be straightforwardly interpreted from the positive semi-definite of the the quantum metric tensor as elaborated in \cite{arxiv2001.05946}.

Our measurements of the quantum geometric tensor via Bloch state tomography in 2D optical Raman lattices exhibit a paradigm to investigate geometry and topology for quantum states.
Since such tomography technique only requires accurate control over the momentum transfer and the phase of the Raman pulse,
it can be readily generalized to interacting systems~\cite{PhysRevA.101.013631}, multi-band systems \cite{EulerClass}, and 3D topological systems~\cite{realization3DSOC}.
It also opens the possibility of spatiotemporal coherent manipulation towards detecting new dynamical topological states of matter.
For instance, people could apply the tomography to quench dynamics with the goal of probing the complete structure of Hopf fibration~\cite{linkingNumber,HopfRamanLattice,HopfFibration,Tarnowski2019}.
The quantum metric tensor also provides important information that the Berry curvature incapable to tell when the Berry curvature vanishes or diverges, including geometric or topological properties~\cite{PhysRevB.103.L241102}, non-Hermatian systems~\cite{PhysRevB.103.125302,Chen2019}, geometric orbital susceptibility~\cite{PhysRevB.94.134423} as well as Bose and Fermi superfluid~\cite{Peotta2015,PhysRevA.97.063625}.
Such situation can be reached by adjusting the parameters of the Raman lattices so that there exist gapless bands and one can detect certain phenomena in topological semi-metals governed by the quantum metric tensor, such as the dynamics of the wave packet
~\cite{QGT3,PhysRevB.103.125302}.

\begin{acknowledgments}
We acknowledge insightful discussions with Ji-Zhou Wu. This work was supported by the Innovation Program for Quantum Science
and Technology (Grant No. 2021ZD0302001 and 2021ZD0302100), the National Natural Science Foundation of China (Grant No. 12025406 and 12104445), Anhui Initiative in Quantum Information Technologies (Grant No. AHY120000), Shanghai Municipal Science and Technology Major Project (Grant No. 2019SHZDZX01), and the Strategic Priority Research Program of Chinese Academy of Science (Grant No. XDB28000000). J.-Y.Z. acknowledges support from the startup grant of University of Science and Technology of China (Grant No. KY2340000152) and the Sponsored by Shanghai Pujiang Program (Grant No. 21PJ1413600).
C.-R.Y. acknowledges support from China Postdoctoral Science Foundation (Grant No. 2021M703112).
\end{acknowledgments}

%

%

\newpage
\onecolumngrid
\renewcommand\thefigure{S\arabic{figure}}
\setcounter{figure}{0}
\renewcommand\theequation{S\arabic{equation}}
\setcounter{equation}{0}
\makeatletter
\newcommand{\rmnum}[1]{\romannumeral #1}
\newcommand{\Rmnum}[1]{\expandafter\@slowromancap\romannumeral #1@}
\makeatother

\newpage

{
\center \bf \large
Supplementary Materials for: Extracting the Quantum Geometric Tensor of an Optical Raman Lattice by Bloch State Tomography\\
\vspace*{0.0cm}
}

\vspace{4ex}

\maketitle
\subsection{The experimental scheme}
\begin{figure*}
\includegraphics[width=0.7\linewidth]{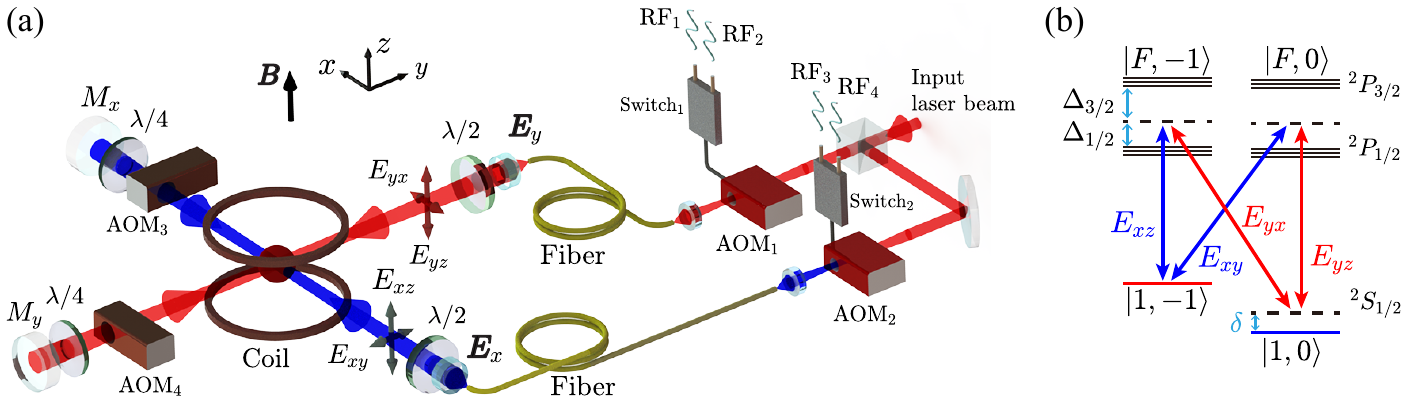}
\caption{(a) The experimental setup.
  The laser beams $\bm{E}_{x,y}$ are used to generate the 2D lattice potential, the Raman potentials and the Raman pulse.
  The frequencies, strengths and phases of $\bm{E}_{x,y}$ are controlled by $\rm{AOM}_{1,2}$, phase-locked radio-frequency signals $\rm{RF}_{1,2,3,4}$ and radio-frequency switches $\rm{Switch}_{1,2}$, respectively.
  $\rm{RF}_{1,2,3,4}$ are generated by two dual-channel arbitrary waveform generators (Keysight 33600A), respectively.
  $\rm{AOM}_{3,4}$ are used to turn off the reflected beams of $\bm{E}_{x,y}$.
  A bias magnetic field $\bm{B} = |\bm{B}| \hat{z}$ with strength $|\bm{B}|\approx 23.4$G along the $\hat{z}$ direction, reducing the quadratic Zeeman splitting of $\sim79$kHz, guarantees the state $\left| F=1, m_F=+1 \right\rangle$ is decoupled for strong Raman strength.
  (b) Level structure and Raman transitions.
  }
\label{FigEsetup}
\end{figure*}

The experimental scheme is shown in Fig.\ref{FigEsetup}.
An input laser beam with wavelength $\lambda=787$nm is split into two sub-lasers by the polarized beam splitter (PBS) and their frequencies are shifted by acousto-optic modulators ($\rm{AOM}_{1,2}$).
The two sub-lasers are transferred to the experimental platform by polarization maintaining fibers, and then pass through a high extinction ratio PBS separately to generate linear polarized laser beams $\bm{E}_{x}$ and $\bm{E}_{y}$.
$\bm{E}_{x,y}$ irradiate the atomic cloud and are reflected back by mirrors $M_{x,y}$.
Meanwhile, $\bm{E}_{x,y}$ are split into two orthogonally polarized components $\bm{E}_{x,y}=\bm{E}_{xy,yx}+\bm{E}_{xz,yz}$ by $\lambda/2$ wave plates and their phases are delayed by $\lambda/4$ wave plates.
The ratio of $\bm{E}_{x,y}$ reflected back to the atomic cloud, marked as $\gamma$, is controlled by $\rm{AOM}_{3,4}$. 
We note that, the polarization of $\bm{E}_{x,y}$ must be not changed with time after the laser passes through $\text{AOM}_{3,4}$, so we use $\text{AOM}_{3,4}$ with an acousto-optical material called {\lq\lq}dense flint glass{\rq\rq}.
Thus, the component of laser beams $\bm{E}_{xy,xz,yx,yz}$ can be written as
\begin{align}\label{laserField}
\begin{split}
{\bm E_{xy}} &=\left|E_{xy}\right|e^{i(\alpha+\alpha_1+\alpha_{L}/2)}\left[ e^{i(k_0x-\alpha_L/2)}+\gamma e^{-i(k_0x-\alpha_L/2)} \right],\\
{\bm E_{xz}}&=\left|E_{xz}\right|e^{i(\alpha+\alpha_1+\alpha_{L}/2)}\left[ e^{i(k_0x-\alpha_L/2)}-\gamma e^{-i(k_0x-\alpha_L/2)} \right], \\
{\bm E_{yx}}&=\left|E_{yx}\right|e^{i(\beta+\beta_1+\beta_{L}/2)}\left[ e^{-i(k_0y+\beta_L/2)}+\gamma e^{i(k_0y+\beta_L/2)} \right],\\
{\bm E_{xz}}&=\left|E_{yz}\right|e^{i(\beta+\beta_1+\beta_{L}/2)}\left[ e^{-i(k_0y+\beta_L/2)}-\gamma e^{i(k_0y+\beta_L/2)} \right],
\end{split}
\end{align}
where the recoil momentum (wave number) $k_0=2\pi/\lambda$, $\left|E_{xy,xz,yx,yz}\right|$ are amplitudes of $\bm{E}_{xy,xz,yx,yz}$, $\alpha_1$($\beta_1$) is the phase controlled by the ratio-frequency (RF) signal input to $\rm{AOM}_{1}$ ($\rm{AOM}_{2}$), $\alpha$ ($\beta$) are initial phases of $\bm{E}_x$ ($\bm{E}_y$), and $\alpha_L(\beta_L)$ are the phases that $\bm{E}_{x}$ ($\bm{E}_{y}$) passes an additional optical path from the atom cloud to mirror $M_x(M_y)$, then back to the atom cloud.
Two magnetic sublevels $\left| F=1, m_F=-1 \right\rangle \equiv \left|\uparrow \right\rangle$ and $\left|F=1, m_F=0 \right\rangle \equiv \left|\downarrow \right\rangle$, split by a bias magnetic field 
$\bm{B} = |\bm{B}| \hat{z}$ with strength $|\bm{B}|\approx 23.4$G
(corresponding to a Zeeman splitting of 16.5MHz) along $\hat{z}$ direction, are coupled by laser beams $\bm{E}_{x,y}$~\cite{W.S_realization2DSOC}.
Thus, the effective Hamiltonian takes the form of Eq.(1) in the main text where the square lattice $V_{\rm{latt}}(x,y)$ and Raman potentials $\Omega_{1,2}(x,y)$ can be written as
\begin{align}
&\Omega_1(x,y)=\frac{\Omega_0}{4}e^{-i(\alpha_1-\beta_1)}\left( e^{-ik_0x}-\gamma e^{ik_0x} \right)\left( e^{-ik_0y}+\gamma e^{ik_0y} \right),\nonumber\\
&\Omega_2(x,y)=\frac{i\Omega_0}{4}e^{-i(\alpha_1-\beta_1)}\left( e^{-ik_0x}+\gamma e^{ik_0x} \right)\left( -e^{-ik_0y}+\gamma e^{ik_0y} \right),\nonumber\\
&V_{\rm{latt}}(x,y)=\gamma V_0\left[ \cos^2(k_0x)+\cos^2(k_0y) \right].
\end{align}
Here, we have considered a unitary transformation using $U=\rm{diag}(e^{i\phi},e^{-i\phi})$ with the global phase $\phi=(\alpha+\alpha_L/2-\beta-\beta_L/2)/2$ together with a coordinate transformation: $k_0x-\alpha_L/2 \mapsto k_0x$, $k_0y-\beta_L/2 \mapsto k_0y$.

\emph{2D Raman lattices.}--For $\gamma=1$, the optical lattices take the form $V_{\rm{latt}}(x,y)=V_0[\cos^2(k_0x)+\cos^2(k_0y)]$ with lattice depth $V_0$, and Raman potentials
take the form $\Omega_{1}(x,y)=\Omega_0\cos(k_0x)\sin(k_0y)$ and $\Omega_2(x,y)=\Omega_0\cos(k_0y)\sin(k_0x)$ with Raman strength $\Omega_0$.
Here, without loss of generality,
we has set $\alpha_1=\beta_1=0$.
It is found that the Raman potentials take a momentum shift of ($\pm k_0, \pm k_0$) between $\left| \uparrow \right\rangle$ and $\left| \downarrow \right\rangle$.
Therefore, the eigen-state $\left| \Psi(\bm{q}) \right\rangle$ can be expanded by a complete basis of plane waves $ |\bm{q}+n\bm{b}_1+l\bm{b}_2 \rangle=e^{i(q_x+2k_0n)x}e^{i(q_y+2k_0l)y}/\sqrt{S}$ and $| \bm{q}+\bm{k}_0+v\bm{b}_1+w\bm{b}_2 \rangle=e^{i(q_x+2k_0v+k_0)x}e^{i(q_y+2k_0w+k_0)y}/\sqrt{S}$~\cite{realization2DSOC}, that is,
\begin{align}\label{wavefunction}
\begin{split}
\left| \Psi(\bm{q}) \right\rangle&=\binom{\sum_{n,l}C_{n,l,\uparrow}|\bm{q}+n\bm{b}_1+l\bm{b}_2 \rangle}{\sum_{v,w}C_{v,w,\downarrow}| \bm{q}+\bm{k}_0+v\bm{b}_1+w\bm{b}_2 \rangle},
\end{split}
\end{align}
where $C_{n,l,\uparrow}$ and $C_{n,l,\downarrow}$ are the complex coefficients, $\bm{b}_{1,2}$
are the vectors in the reciprocal lattice, $(n,l,v,w)$ are integers, and $S$ denotes the system area.
If only quasi-momentum in the FBZ is considered, the eigen-state can be simply expressed as $\left| \Psi(\bm{q}) \right\rangle=c_{\uparrow}(\bm{q})\left|\bm{q},\uparrow \right\rangle+c_{\downarrow}(\bm{q})\left|\bm{q}+\bm{k}_0,\downarrow \right\rangle$.

\emph{Raman pulse}--For $\gamma=0$, only the incident beams of $\bm{E}_{x,y}$ are retained, which is used to generate the Raman pulse.
The optical lattices vanish.
Raman potentials turns into the Raman coupling $\Omega_{\text{R}}=\Omega_{\text{R0}}e^{-i[k_0(x+y)-\Delta\varphi]}$ with Raman pulse strength $\Omega_{\text{R0}}=\sqrt{2}\Omega_0/4$, which takes a momentum shift of ($-k_0,-k_0$) between $\left| \uparrow \right\rangle$ and $\left| \downarrow \right\rangle$.
The relative phase between the Raman pulse and the Raman lattices $\Delta\varphi=-\pi/4-\alpha_1+\beta_1$ only depends on the
phases $\alpha_1$ and $\beta_1$.
The Hamiltonian of Raman pulse is then given by
\begin{equation} \label{Eq:RamanPulse}
H_{\rm{R}}=\begin{pmatrix}
  \bm{k}^2/2m+\delta_{\text{R}}/2& \Omega_{\text{R}}\\
  \Omega_{\text{R}}^*&\bm{k}^2/2m-\delta_{\text{R}}/2
\end{pmatrix}.
\end{equation}

\subsection{Bloch state tomography in Raman lattices}
The purpose of the tomography is to measure the expectation values of three Pauli matrices $\langle\sigma_{i=x,y,z}\rangle$, which can be written as
\begin{align}\label{QAHsigmaxyz}
\begin{split}
&\left\langle \sigma_x \right\rangle=\sum_{n,l}\left( C_{n,l,\uparrow}^*C_{n,l,\downarrow}+C_{n,l,\uparrow}C_{n,l,\downarrow}^* \right),\\
&\left\langle \sigma_y \right\rangle=\sum_{n,l}\left( -iC_{n,l,\uparrow}^*C_{n,l,\downarrow}+iC_{n,l,\uparrow}C_{n,l,\downarrow}^* \right),\\
&\left\langle \sigma_z \right\rangle=\sum_{n,l}\left( \left| C_{n,l,\uparrow} \right|^2-\left| C_{n,l,\downarrow} \right|^2 \right),
\end{split}
\end{align}
according to Eq.(\ref{wavefunction}).
We apply a momentum-transferred Raman pulse to measure $\langle \sigma_{x,y,z} \rangle$.
When the Raman pulse is applied to the eigen-state of Raman lattices $\left| \Psi \right\rangle$ with a duration time $t_{\text{R}}$, the state evolves unitarily to a new state
\begin{equation}\label{EqNewState}
| \widetilde{\Psi}(\bm{q},t_{\text{R}}) \rangle=\exp(-iH_{\text{R}}t_{\text{R}})\left| \Psi \right\rangle
\end{equation}
After spin-resolved ToF image, the spin polarization $P=\left\langle \widetilde{\Psi} \left| \sigma_z \right| \widetilde{\Psi} \right\rangle$ is obtained.
In order to get a concise expression for the physical process above, we neglect the kinetic energy term $\bm{k}^2/2m$ (as is justified below) and set $\delta_{\text{R}}=0$.
The Hamiltonian in Eq.~(\ref{Eq:RamanPulse}) then takes a simplified off-diagonal form ${H}_{\rm{R}}=\Omega_{\text{R0}}\cos[k_0(x+y)-\Delta\varphi]\sigma_x+\Omega_{\text{R0}}\sin[k_0(x+y)-\Delta\varphi]\sigma_y$.
Thus, the spin polarization can be expressed as
\begin{align}\label{expectedValue}
\begin{split}
P&=( | c_{\uparrow} |^2-| c_{\downarrow} |^2 )\cos(2\Omega_{\text{R0}}t_{\text{R}})\\
&+\left( -ie^{i\Delta\varphi}c_{\uparrow}c_{\downarrow}^* \langle \bm{q} | e^{-ik_0(x+y)} | \bm{q}+\bm{k_0} \rangle +ie^{-i\Delta\varphi}c_{\uparrow}^*c_{\downarrow}\langle \bm{q}+\bm{k}_0 | e^{ik_0(x+y)} | \bm{q} \rangle \right)\sin(2\Omega_{\text{R0}}t_{\text{R}}).
\end{split}
\end{align}
Using the relation $\langle \bm{q}+n\bm{b}_1+l\bm{b}_2 | e^{-ik_0(x+y)} | \bm{q}+\bm{k_0}+v\bm{b}_1+w\bm{b}_2 \rangle=\delta_{n,v}\delta_{l,w}$ and $\langle \bm{q}+\bm{k}_0+v\bm{b}_1+w\bm{b}_2 | e^{ik_0(x+y)} | \bm{q}+n\bm{b}_1+l\bm{b}_2 \rangle=\delta_{n,v}\delta_{l,w}$, one 
finds that the momentum transferred by Raman lattices is counteracted by the Raman pulse.
Then, combining Eq.(\ref{QAHsigmaxyz}), Eq.(\ref{expectedValue}) can be written as the superposition of $\left\langle\sigma_{x,y,z}\right\rangle$, as shown by Eq.~(3) in the main text.
Equation.~(3) is actually a sinusoidal function that evolves with $t_{\text{R}}$ and possesses an amplitude of the oscillations $\text{Amp}=\sqrt{(\langle \sigma_{z} \rangle)^2+(\langle \sigma_y \rangle \cos\Delta\varphi+\langle \sigma_x \rangle\sin \Delta\varphi)^2}$.
Therefore, $\left\langle\sigma_{x,y,z}\right\rangle$ are extracted by tuning $\Delta\varphi$ and $t_{\text{R}}$.
When $t_{\text{R}}=0$, $\left\langle\sigma_{z}\right\rangle$ is obtained.
When $t_{\text{R}}=\pi/(4\Omega_{\text{R0}})$, $\left\langle\sigma_{x}\right\rangle$ ($\left\langle\sigma_{y}\right\rangle$) is obtained via a $\pi/2$ Raman pulse with $\Delta\varphi=\pi/2 (0)$.

\begin{figure}
  \center
  \includegraphics[width=0.6\linewidth]{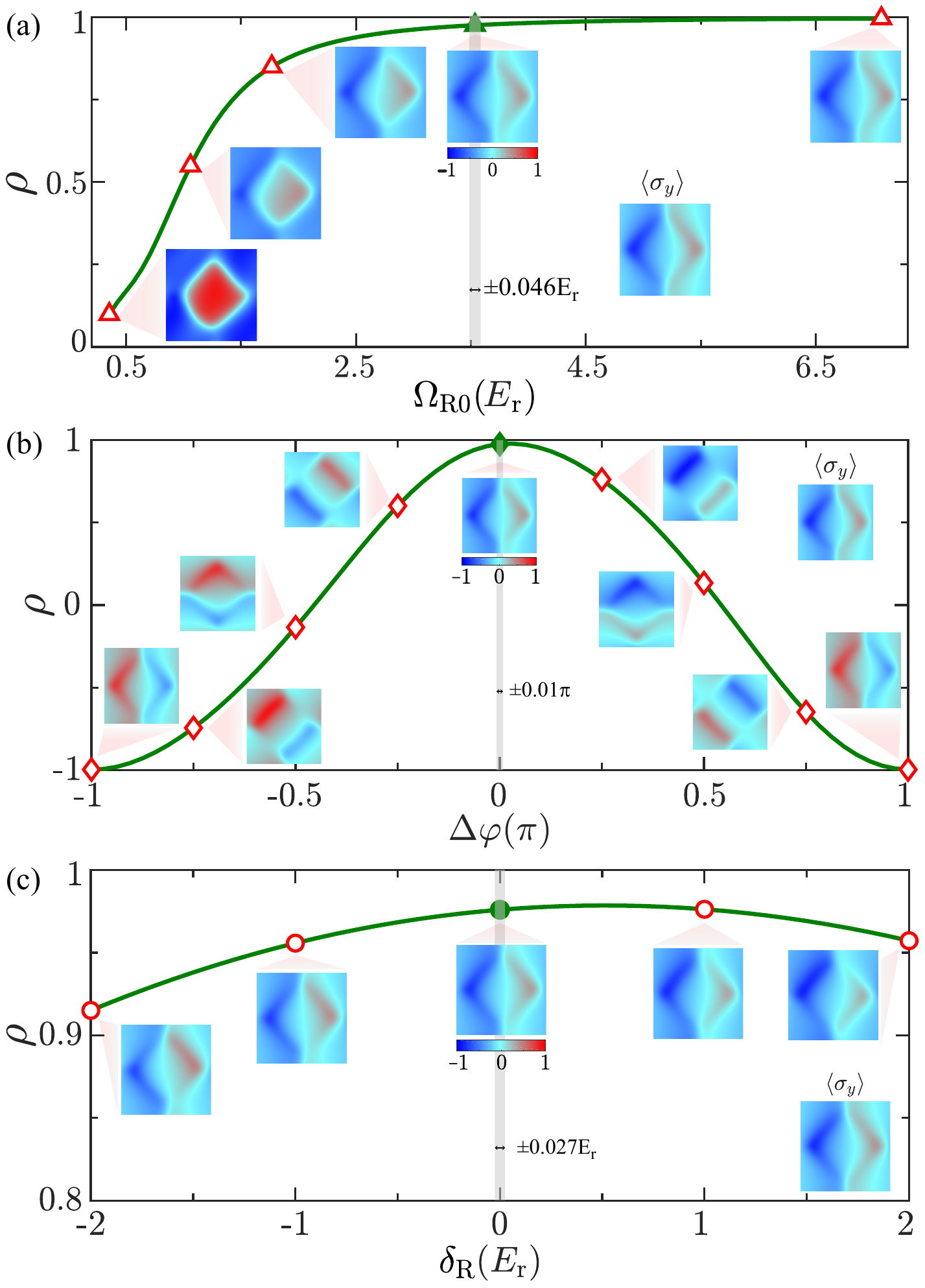}
 \caption{The sensitivity of the Bloch state tomography.
 Correlation coefficient $\rho$ between $P(\bm{q},t)$ and $\left\langle\sigma_y(\bm{q},t)\right\rangle$ for the lowest band as a function of $\Omega_{\rm{R0}}$ (a), $\Delta\varphi$ (b) and $\delta_{\rm{R}}$ (c), respectively.
The solid curves represent $\rho$ versus $\Omega_{\rm{R0}}$, $\Delta\varphi$ and $\delta_{\rm{R}}$.
Insets: the typical $P$ in FBZ is shown and their location is marked by triangles, diamonds and circles.
The parameters of the green triangle, diamond and circle are $\Omega_{\rm{R0}} \approx 3.4E_{\rm{r}}$, $\delta_{\rm{R}}=0$ and $\Delta\varphi=\pi/2$, and the corresponding correlation coefficient $\rho>0.97$.
The small shaded gray regions indicate the measurement errors of $\Omega_{\rm{R0}}$, $\Delta\varphi$ and $\delta_{\rm{R}}$, which are from multiple experimental observations.
Other results in the figure are the numerical simulations.
Parameters: $(V_0,\Omega_0,\delta)=(4,1,0.2)E_{\rm{r}}$;
(a) $\Delta\varphi=0$ and $\delta_{\rm{R}}=0$, (b) $\delta_{\rm{R}}=0$ and $\Omega_{\rm{R0}}\approx3.4E_{\rm{r}}$, (c) $\Delta\varphi=0$ and $\Omega_{\rm{R0}}\approx3.4E_{\rm{r}}$.}
  \label{FigCorr}
\end{figure}

Next, we estimate the impact of the kinetic energy term in the Raman pulse on the measurement of $\langle \sigma_{x,y} \rangle$.
To this end, we firstly calculate numerically the spin polarization $P$ for the lowest band.
In the numerical calculation, we set $\gamma=10\%$.
Then, the correlation coefficient $\rho$ between $P$ and $\langle\sigma_{x,y}\rangle$ as a function of $\Omega_{\text{R0}}$ is calculated to evaluate under how strong Raman pulse can we ignore the kinetic energy term.
The correlation coefficient is written as $\rho(P,\langle\sigma_{i}\rangle)=\text{cov}(P,\langle\sigma_{i}\rangle)/{\sigma_{P}\sigma_{\langle\sigma_{i}\rangle}}$, where $\text{cov}(P,\langle\sigma_{i}\rangle)$ and $\sigma_{P,\langle\sigma_{i}\rangle}$ are the covariance and the standard deviation, respectively.
For $\rho=0$, $P$ and $\langle \sigma_{i} \rangle$ are completely uncorrelated; For $\rho=1$, $P$ is exactly same as $\langle \sigma_{i} \rangle$.
As a example, the correlation coefficient $\rho$ versus the strength of Raman pulse $\Omega_{\text{R0}}$ for $\Delta\varphi=0$ is shown in Fig.\ref{FigCorr}(a).
$\rho$ increases from 
$\sim0$ to 1 when $\Omega_{\text{R0}}$ is tuned from $\sim0$ to $7E_{\rm{r}}$.
Meanwhile, the patterns $P(\bm{q},t_{\text{R}})$ become more and more consistent with $\left\langle\sigma_y\right\rangle$ when increasing $\Omega_{\text{R0}}$.
In experiment, we select $\Omega_{\text{R0}}\approx 3.4E_{\rm{r}}$ to realize Raman pulse, which is enough to rotate $\left\langle\sigma_z\right\rangle$ to $\left\langle \sigma_{x,y} \right\rangle$ with $\rho \approx 0.97$.
On the other hand, intuitively, the duration time of $\pi/2$ Raman pulse $t_{\text{R}}=10\mu s$ is much smaller than characteristic time of Raman lattices $t_{\text{c}}=h/E_{\rm{r}}\approx 270\mu s$, and the atoms could hardly evolve
within the time $t_{\text{R}}$.

\subsection{The sensitivities of the Bloch-state tomography}
The sensitivities of the Bloch-state tomography are determined by the parameters associated with the Raman pulse, including the strength $\Omega_{\rm{R0}}$ and the detuning $\delta_{\rm{R}}$ of the Raman pulse, as well as the relative phase between the Raman lattice and the Raman pulse $\Delta\varphi$.
We estimate the effects of $\Omega_{\rm R0}$, $\delta_{\rm{R}}$ and $\Delta\varphi$ when measuring $\langle\sigma_{x,y}(\bm{q})\rangle$ via calculating the correlation coefficient $\rho$ between $P(\bm{q},t_{\rm{R}})$ and $\langle \sigma_{x,y}(\bm{q}) \rangle$ for the lowest band, respectively.
Figure.~\ref{FigCorr} shows an example for $\langle \sigma_y(\bm{q}) \rangle$.
$\rho$ increases from $\sim0$ to 1 as $\Omega_{\rm R0}$ is tuned from $\sim0$ to 7$E_{\rm{r}}$ (Fig.~\ref{FigCorr}(a)).
As long as $\Omega_{\rm R0}$ is strong enough, $P(\bm{q},t_{\rm{R}})$ basically does not change with $\Omega_{\rm R0}$.
As $\Delta\varphi$ tunes, $\rho$ oscillates with the amplitude of 1 (Fig.~\ref{FigCorr}(b)).
$P$ is insensitive to changes in $\Delta\varphi$ close to $\pi/2$.
$P$ changes only slightly when $\delta_{\rm R}$ is adjusted from $-2E_{\rm r}$ to $2E_{\rm r}$ (Fig.~\ref{FigCorr}(c)).
In short, $P$ is rather insensitive to the fluctuation of the parameters, which is close to the selected parameters by this experiment, i.e., $\Omega_{\rm{R0}} \approx 3.4E_{\rm{r}}$, $\delta_{\rm{R}}=0$ and $\Delta\varphi=\pi/2$.

In experiment, we also measure the standard deviation of these three parameters.
The standard deviation of $\Omega_{\rm R0}$, $\delta_{\rm{R}}$ and $\Delta\varphi$ are respectively $\pm 0.046E_{\rm r}$, $\pm 0.01\pi$ and $\pm 0.027E_{\rm r}$ (gray shaded regions in Fig.~\ref{FigCorr}).
For all three cases, $\rho>0.97$ within the standard deviation.
Therefore, the Raman pulse is very stable.
In other words, the Bloch-state tomography is insensitive to the perturbation of the present parameters.

\begin{figure}
  \center
  \includegraphics[width=0.6\linewidth]{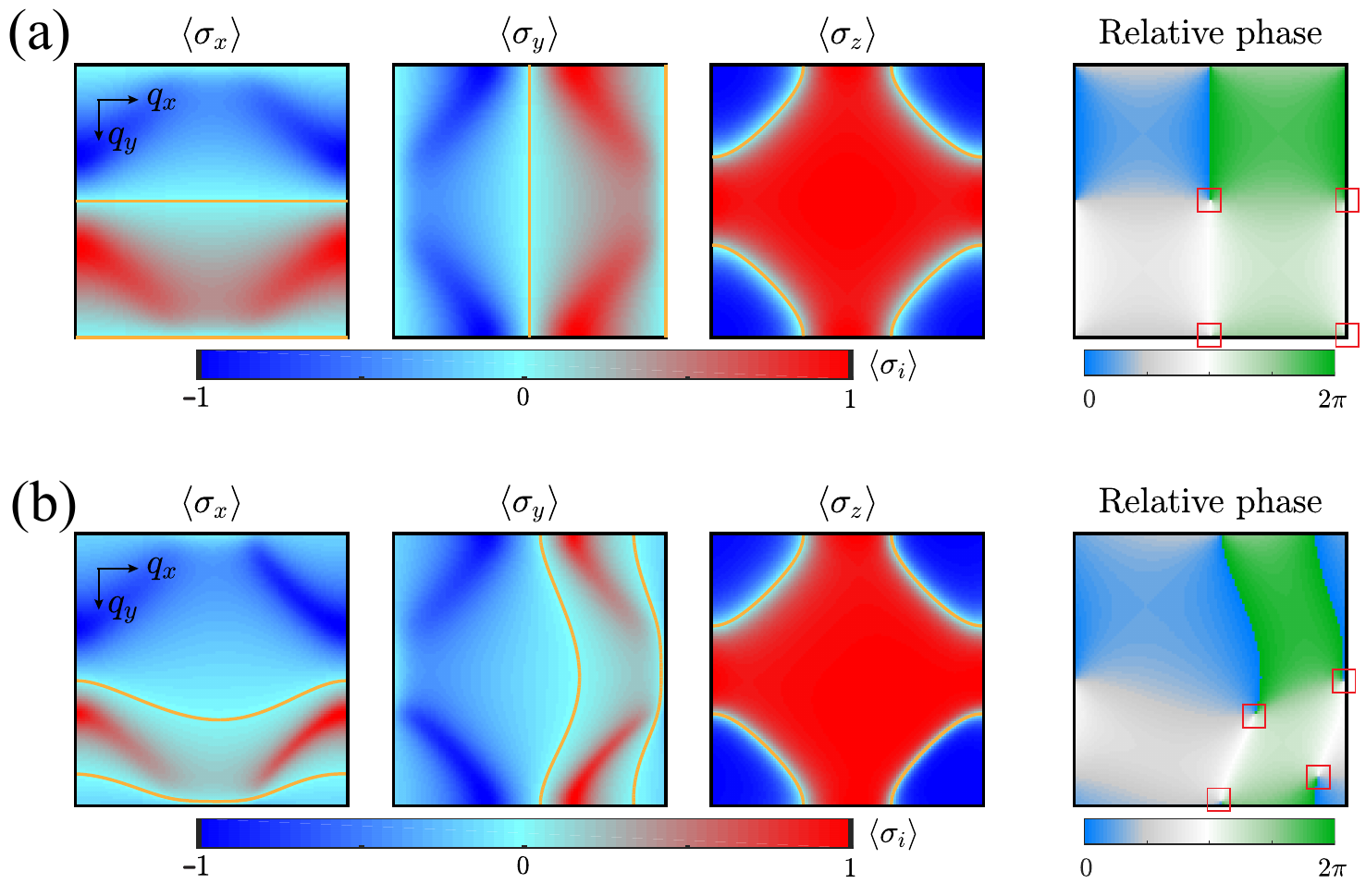}
 \caption{The expectation value of three Pauli matrices $\langle \sigma_{x,y,z} \rangle$ in the lowest band and the relative phase between the eigenfunction $u_{\uparrow,\downarrow}$ under two-band tight-binding approximation of the Hamiltonian Eq.(1) (a) and the multiple bands determined by Eq. (1) (b).
 The yellow solid curves are determined by $\langle\sigma_{x,y,z}\rangle=0$.
 The intersections of the yellow curves in $\langle\sigma_{x,y}\rangle$
 locate the phase vortices, which are marked by red squares.
 Here, we emphasize that the normalization for $P_{x,y,z}$ does not change the relative phase between the eigenfunction $u_{\uparrow,\downarrow}$ in the FBZ.
 Parameters: $(V_0,\Omega_0,\delta)=(4,1,-0.2)E_{\rm{r}}$; $t_0\approx 0.09\propto V_0$, $t_{\rm{so}}\approx -0.05\propto\Omega_0$.
}
  \label{FigSigmaxyzPsiPhase}
\end{figure}
\subsection{The origin of the phase vortices deviating from the high-symmetry points}
The asymmetry in the phase vortices in Fig.3 stems mainly from the effects of the higher bands, since the eigenfunction of the lowest band is modified by the higher bands.
We calculate the expectation values of three Pauli matrices by both the two-bands tight-binding approximation
(TBA) of the Eq.~(1) (see below) and the Eq.~(1) with multiple-bands.
For simplicity, $\langle \sigma_{x,y,z} \rangle$ here refer to the results evaluated in the lowest band.
(a) The Bloch Hamiltonian within TBA is $\mathcal{H}_{\rm TBA}=\bm{h} \cdot \bm{\sigma}=h_x\sigma_x+h_y\sigma_y+h_z\sigma_z=2t_{\rm so}\sin {q_y}\sigma_x+2t_{\rm so}\sin {q_x}\sigma_y+(\delta/2-2t_{\rm so}(\cos q_x+\cos q_y))\sigma_z$, which gives $\langle \sigma_{x,y,z} \rangle= - h_{x,y,z}/|\bm{h}|$.
(b) $\langle \sigma_{x,y,z} \rangle$ of the Hamiltonian Eq.(1) with the multiple-bands are obtained by the formulas $\langle \sigma_{x,y,z} \rangle=P_{x,y,z}/\sqrt{P_x^2+P_y^2+P_z^2}$ and $P_{x,y,z}=\langle \Psi(\bm{q})|\sigma_{x,y,z}|\Psi(\bm{q}) \rangle$, where $|\Psi(\bm{q}) \rangle$ is the eigenfunction of the Hamiltonian Eq.(1).
Here, the eigenfunction of the lowest band is modified by the coupling between the lowest band and the higher bands.
We normalize $P_{x,y,z}$ for each band in order to calculate the eigenfunctions, the Berry curvature and the quantum metric tensor of the lowest bands.
Basically, such normalization is reasonable for the present parameters and hardly changes the symmetry of $P$ (see Fig.~\ref{FigCorr} and Fig.~\ref{FigSigmaxyzPsiPhase} for $\langle \sigma_{y} \rangle$).

In Fig.~\ref{FigSigmaxyzPsiPhase}, we plot the expectation values of the three Pauli matrices $\langle \sigma_{x,y,z} \rangle$ and the relative phase between the eigenfunction $u_{\uparrow,\downarrow}$ with the lowest band for the above two cases.
The relative phase between the eigenfunction is obtained by $\langle \sigma_{x} \rangle$ and $\langle \sigma_{y} \rangle$, i.e., $\phi_{\bm{q}}=\arg(\left\langle \sigma_x \right\rangle+i\left\langle \sigma_y \right\rangle)\in[0,2\pi)$.
$\langle \sigma_{z} \rangle$ ($\langle \sigma_{x,y} \rangle$) is symmetric along the $q_x$ and $q_y$ ($q_{x,y}$) directions within the TBA of Eq.~(1), being consistent with the multiple-band results determined by Eq.~(1).
However, there also exist some subtle differences.
Figure~\ref{FigSigmaxyzPsiPhase}(a) shows two straight lines in $\langle \sigma_{x,y} \rangle=0$ along the $q_{x,y}$ directions, whereas these two straight lines are bent due to the effects of the higher bands, shown in Fig.~\ref{FigSigmaxyzPsiPhase}(b).
The phase vortices are determined by $(\langle \sigma_x \rangle,\langle \sigma_y \rangle,\langle \sigma_z \rangle)=(0,0,\pm1)$, whose locations are indicated by the intersections of the yellow straight or bent lines in the FBZ.
We then see that, due to the effects of the higher bands, the position of the phase vortices deviates from the high-symmetry points.

%

\subsection{The experimental protocol}
We measure the expectation values of three Pauli matrices $\langle \sigma_{x,y,z} \rangle$ by applying a Raman pulse to the eigen-state of Raman lattices.
We divide the experimental protocol into three steps: 

\begin{figure}
  \center
  \includegraphics[width=1\linewidth]{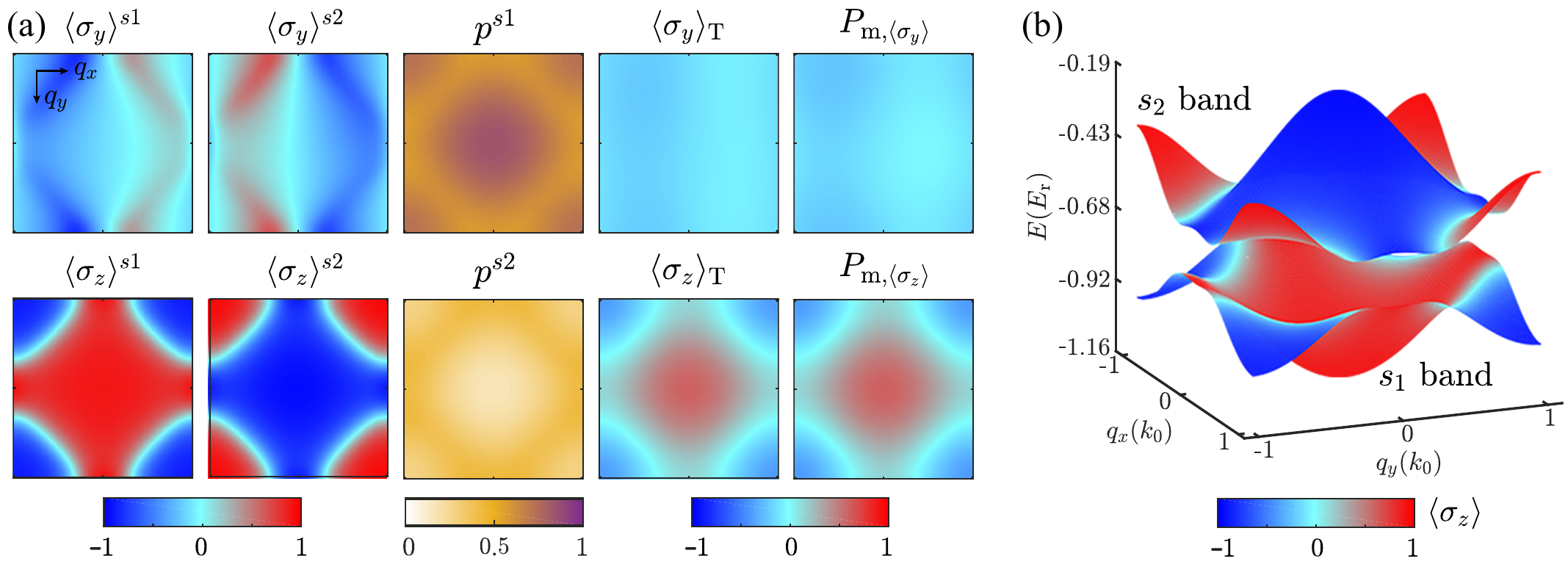}
 \caption{(a) The expectation value of Pauli matrices $\langle\sigma_{y,z}\rangle^{s1,s2}$ and the ratio of the atoms $p^{s1,s2}$ in the FBZ.
 $\langle\sigma_{y,z}\rangle_{\rm T}$ and the calculated spin textures $P_{\text{m},\langle \sigma_{y,z} \rangle}$ is obtained by Eq.~\ref{Eqsigma} and Eq.~\ref{EqPm}, respectively.
 (b) The lowest two bands $s1$ and $s2$.
 Parameters: $(V_0,\Omega_0,\delta)=(4,1,-0.2)E_{\rm{r}}, t_{\rm{R}}=10\mu s,\delta_{\rm{R}}=0, \Omega_{\rm{R0}}\approx 3.4E_{\rm{r}}$.
}
  \label{FigAtomPopulationHighBand}
\end{figure}

\emph{{\romannumeral1}. The preparation of the state}--The atoms is cooled by evaporative cooling.
Whereafter, the atoms are loaded adiabatically into Raman lattices in 100ms by ramping the intensity of $\bm{E}_{x,y}$ with the temperature $\sim100\text{nK}$, which is controlled by the power of $\rm{RF}_{1,3}$.
Meanwhile, the phase and frequency of $\rm{RF}_{1,3}$ are fixed.
And $\rm{AOM}_{3,4}$ and $\rm{RF}_{2,4}$ are turned off.
Then, the atoms occupy each momentum point in the Brillouin zone; most of them are in the lowest band and a small part of them are in the higher band.

\emph{{\romannumeral2}. The realization of the Raman pulse}--The Raman pulse is realized by simultaneously performing
4 manipulations on the beams $\bm{E}_{x,y}$:
1) The reflected beams of $\bm{E}_{x,y}$ are shut down by turning on $\rm{AOM}_{3,4}$.
The ratio of beams $\bm{E}_{x,y}$ reflected back the atoms $\gamma$ is about 10$\%$, which is evaluated by calibrating the optical lattice depth with $\rm{AOM}_{3,4}$ turned on.
The reason for $\gamma>0$ is that the diffraction efficiency of AOM can not reach 100$\%$.
2) The strength of the Raman pulse is set as $\Omega_{\text{R0}}\approx3.4E_{\rm{r}}$ by tuning the intensity of $\bm{E}_{x,y}$.
3) The relative phase $\Delta\varphi$ is adjusted by tuning the phase of $\bm{E}_{x,y}$.
$\Delta\varphi$ is determined by matching the experimental measurements and numerical calculations [In fact, the absolute $\Delta\varphi$ is not necessary. In experiment, only a difference of $\Delta\varphi=\pi/2$ is necessary for measuring $\langle\sigma_x\rangle$ and $\langle\sigma_y\rangle$. In order to compare the experiment and theory at the same $\Delta\varphi$, we make the phase of the experimental data consistent with the numerical calculation by shifting a offset of the experimental $\Delta\varphi$.], as depicted in Fig.1(e).
And the stability of $\Delta\varphi$ is assessed by comparing the spin polarization $P_{\rm{m}}$ in the presence and absence of the Raman pulse.
The experimental measurements show that the uncertainty of $P_{\rm{m}}$ are both 0.03 for 100 photos regardless of the presence of the Raman pulse, which indicates that the uncertainty of $\Delta\varphi$ does not disturb the experimental observations.
4) The detuning of Raman pulse $\delta_{\rm{R}}$ is set as 0 by adjusting the frequency of $\bm{E}_{x,y}$.
Meanwhile, ratio-frequency signals are switched from $\rm{RF}_{1,3}$ to $\rm{RF}_{2,4}$ by $\rm{Switch}_{1,2}$, and the switching speed is about 200ns, which is much less than $t_{\text{R}}=10\mu$s.
The intensity, phase and frequency of $\bm{E}_{x,y}$ are exactly controlled by the power, phase and frequency of $\rm{RF}_{1,2,3,4}$ input to $\rm{AOM}_{1,2}$, respectively.

\emph{{\romannumeral3}. The detection of the spin polarization}--After the Raman pulse is applied to the atoms for a certain time $t_{\text{R}}$, 
all the lasers and the bias field are switched off in less than $1\mu$s.
Then, the Stern-Gerlach magnetic field is turned on so that the atom cloud of $| \uparrow \rangle$ and $| \downarrow \rangle$ are separated.
After the atoms expand freely for 25ms, we take photos of the atoms along $\hat{z}$ direction to obtain the atomic distribution in $| \uparrow \rangle$ and $| \downarrow \rangle$.
Thus, the atom numbers $n_{\uparrow}(\bm{q})$ and $n_{\downarrow}(\bm{q})$ are obtained.

\begin{figure}
  \includegraphics[width=0.7\linewidth]{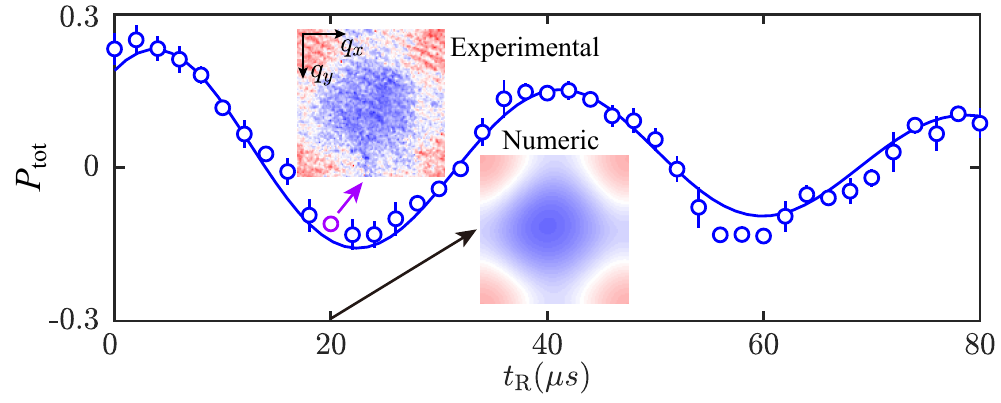}
  \caption{The spin polarization as a function of the duration time $t_{\rm{R}}$ of Raman pulse with $T=100$nK.
  The dots with error bars (solid curve) is experimental data (the fitting using damped sine function).
  The spin textures are from the original experimental measurements and numerical calculations at $t_{\rm{R}}=20\mu s$.
  Parameters: $(V_0,~\Omega_0,~\delta)=(4.0,1.0,-0.2)E_{\rm{r}}$, $\Delta\varphi=\pi,~\Omega_{\rm{R0}}\approx 3.4E_{\rm{r}}$.
  }
  \label{FigCoherentAnalyse}
\end{figure}

\subsection{Data analysis}
We first analysis the experimental protocol measuring the spin textures.

i. The atoms are loaded into Raman lattices in 100ms, which can be regarded as an adiabatic process and then induces the populations of the atom in accordance with the Bose distribution.
The ratio of the atoms populated to the momentum point $\bm{q}$ of the $j$th band $p^{j}(\bm{q})\in [0,1]$ is calculated numerically by the Bose distribution with the input parameters, including atomic temperature $T$, atomic density $n_d$ and the band structures, as shown in Fig.~\ref{FigAtomPopulationHighBand}.
Here, $T$ and $n_d$ are measured experimentally, and the band structures are obtained by diagonalizing the Hamiltonian Eq.(1) in the main text.
The atoms are almost entirely distributed in the two lowest $s$ bands when the atomic temperature is $T=100$nK and atomic density is $n_d=3\times 10^{18}m^{-3}$.
About 67\% (33\%) atoms populate to the $s1~(s2)$ band for $\delta=-0.2E_{\rm{r}}$.

ii. After applying the Raman pulse, the spin polarization evolves according to by Eq. (3) in the main text.
Numerically, the measured spin textures after applying a Raman pulse can be decomposed into the sum of the individual bands, i.e.,
\begin{align}\label{EqPm}
P_{\rm m}(\bm{q})=\frac{n_{\uparrow}(\bm{q})-n_{\downarrow}(\bm{q})}{n_{\uparrow}(\bm{q})+n_{\downarrow}(\bm{q})}=\frac{\sum_{j=1}^{j=N} p^{j}(\bm{q})[A_{\uparrow}^{j}(\bm{q})-A_{\downarrow}^{j}(\bm{q})]}{\sum_{j=1}^{j=N} p^{j}(\bm{q})[A_{\uparrow}^{j}(\bm{q})+A_{\downarrow}^{j}(\bm{q})]},
\end{align}
where $A_{\uparrow}^{j}=|\widetilde{\Psi}_{\uparrow}^{j}(\bm{q})|^2$ ($A_{\downarrow}^{j}=|\widetilde{\Psi}_{\downarrow}^{j}(\bm{q})|^2$) is determined by the evolved state $|\widetilde{\Psi}_{\uparrow}^{j}(\bm{q})\rangle$ ($|\widetilde{\Psi}_{\downarrow}^{j}(\bm{q})\rangle$) of $\mid\uparrow\rangle$ ($\mid\downarrow\rangle$) in the $j$th band.
 Here, we have assumed that $p^{j}(\bm{q})$ does not change over time with $t_{\rm{R}}\in [0, 10)\mu s$, which is reasonable according to the following numerical evidence and experimental data.

For the numerical evidence, we can calculate the expectation values of three Pauli matrices with non-zero temperature by the sum of $\langle \sigma_{x,y,z} \rangle^{j}$ with the ratio $p^{j}(\bm{q})$, i.e.,
\begin{equation}\label{Eqsigma}
\langle \sigma_{x,y,z}(\bm{q}) \rangle_{\rm T}=\sum_{j=1}^{j=N}p^{j}(\bm{q})\langle \sigma_{x,y,z}(\bm{q}) \rangle^{j}.
\end{equation}
From Eq.~(\ref{Eqsigma}), $\langle \sigma_{x,y,z}(\bm{q}) \rangle_{\rm T}$ contains the corrections of higher-band population.
To validate our assumptions, we need to verify that the expectation value of Pauli matrices measured using Raman pulses is consistent with directly calculated using the eigenstate, i.e., $P_{\text{m},i}\simeq\langle \sigma_{i} \rangle_{\rm T}~(i=x,y,z)$.
Hence, we calculate numerically Eq.~(\ref{Eqsigma}) and Eq.~(\ref{EqPm}) with the atomic temperature $T=100$nK and atomic density $n_d=3\times 10^{18}m^{-3}$, as shown in Fig.~\ref{FigAtomPopulationHighBand}.
For individual bands, the spin textures $\langle\sigma_{y,z}\rangle^{s1,s2}$ have high contrast.
However, about 33\% of the atoms populate the s2 band, which reduces the contrast of the spin textures.
The correlation coefficient $\rho$ between $P_{\text{m},\langle \sigma_{i} \rangle}$ and $\langle\sigma_{i}\rangle_{\rm T}$ is more than 0.9 for a $\pi/2$ Raman pulse of 10$\mu$s, which indicates $P_{\text{m},i}\simeq\langle \sigma_{i} \rangle_{\rm T}$.

For the experimental data, we compare the band gap and the Rabi frequency of the Raman pulse to validate our assumptions.
The band gap is from a few hundred hertz to a few kilohertz depending on which quasimomentum point the atom is located, which is much less (more) than the Rabi frequency of Raman pulse 25kHz (See Eq.~(3) in the main text).
Hence, the atoms moving to other bands should not be observed experimentally.
To verify it, we measure $P_{\rm{tot}}=(N_{\uparrow}-N_{\downarrow})/(N_{\uparrow}+N_{\downarrow})$ with the atom number of $|\uparrow,\downarrow\rangle$ $N_{\uparrow,\downarrow}$, as shown in Fig.~\ref{FigCoherentAnalyse}.
The period of $P_{\rm{tot}}(t_{\rm{R}})$ is about 35$\mu s$, which is determined by the strength $\Omega_{\rm{R0}}$ and has nothing to do with the gap.
Further, the spin texture at $t_{\rm{R}}=20\mu s$ is consistent with the numerical calculations, implying the atoms moving to other bands can be ignored when $t_{\rm{R}}=10\mu s$.

\begin{figure*}
\includegraphics[width=0.8\linewidth]{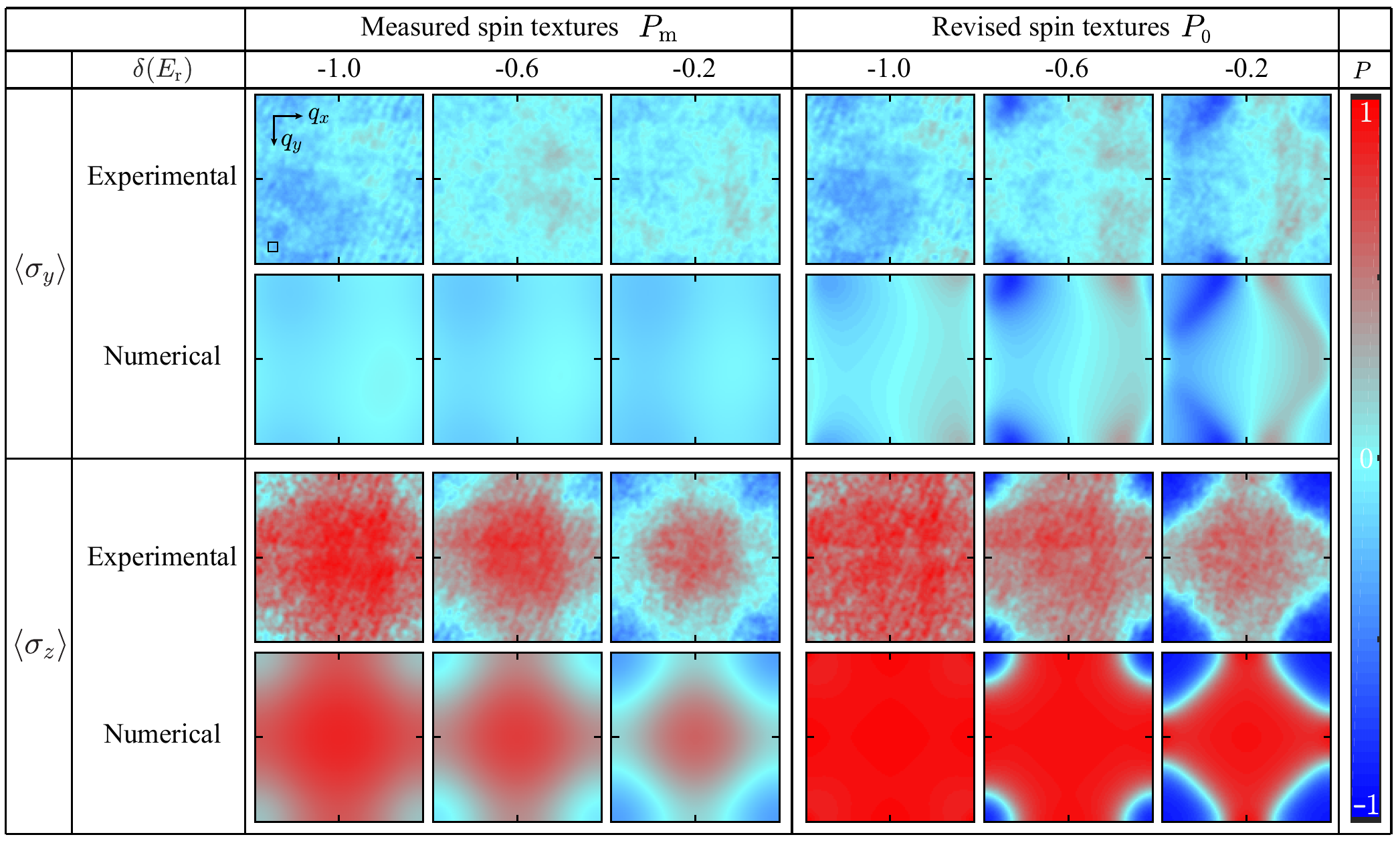}
\caption{
The measured spin textures and the revised spin textures in the lowest band.
Parameters: $(V_0,\Omega_0)=(4,1)E_{\rm{r}}$.
}
\label{FigoriDataToRevisedSpinTexture}
\end{figure*}

Now, we remove the higher band population in the measured spin textures.
The higher band population is removed to get the spin polarization of the lowest band, that is, $P_0(\bm{q})=(n_{\uparrow}^{s1}(\bm{q})-n_{\downarrow}^{s1}(\bm{q}))/(n_{\uparrow}^{s1}(\bm{q})+n_{\downarrow}^{s1}(\bm{q}))$ with atom number of the lowest band for spin up $n_{\uparrow}^{s1}(\bm{q})$ and spin down $n_{\downarrow}^{s1}(\bm{q})$.
To this end, The proportion of the atoms in the lowest band $\eta_{\uparrow,\downarrow}(T)$ needs to be obtained.
According to the evolved states $|\widetilde{\Psi}_{\uparrow}^{j}(\bm{q})\rangle$ ($|\widetilde{\Psi}_{\downarrow}^{j}(\bm{q})\rangle$) of $\mid\uparrow\rangle$ ($\mid\downarrow\rangle$) in the $j$th band calculating by Eq.~(S6) at $t_{\rm{R}}=10\mu s$,
$n_{\uparrow}^{j}(\bm{q})=p^{j}A_{\uparrow}^{j}$ ($n_{\downarrow}^{j}(\bm{q})=p^{j}A_{\downarrow}^{j}$) with $A_{\uparrow}^{j}=|\widetilde{\Psi}_{\uparrow}^{j}(\bm{q})|^2$ ($A_{\downarrow}^{j}=|\widetilde{\Psi}_{\downarrow}^{j}(\bm{q})|^2$).
We then get the proportions $\eta_{\uparrow,\downarrow}(T)=n_{\uparrow,\downarrow}^{s1}/n_{\uparrow,\downarrow}$, where $n_{\uparrow,\downarrow}$ are obtained by experimental measurements.
Thus, the spin polarization of the lowest band is expressed as $P_0(\bm{q})=[\eta_{\uparrow}(T)n_{\uparrow}-\eta_{\downarrow}(T)n_{\downarrow}]/[\eta_{\uparrow}(T)n_{\uparrow}+\eta_{\downarrow}(T)n_{\downarrow}]$.
Whereafter, we apply the correction to the measured data and the revised spin textures $\langle \sigma_{x,y,z} \rangle$ are drawn in Fig.\ref{FigoriDataToRevisedSpinTexture}.
Finally, $\langle \sigma_{x,y,z} \rangle$ are displayed in the main text after 
normalization $\sqrt{\langle \sigma_x \rangle^2+\langle \sigma_y \rangle^2+\langle \sigma_z \rangle^2}=1$ and smoothed between adjacent momentum points.
The smoothing is achieved by calculating the average of the spin polarizations within a square area.
For example, the smoothing of the center point in the black square of Fig.~\ref{FigoriDataToRevisedSpinTexture} is obtained by averaging the spin polarizations in this black square.
The area of the black square is about 0.3\% of the area of the FBZ.
The FBZ with around 12000 momentum points has a high resolution in momentum space and each momentum point is about 2.6$\mu$m, given by the pixel size of the camera.

\begin{figure}
  \center
  \includegraphics[width=0.4\linewidth]{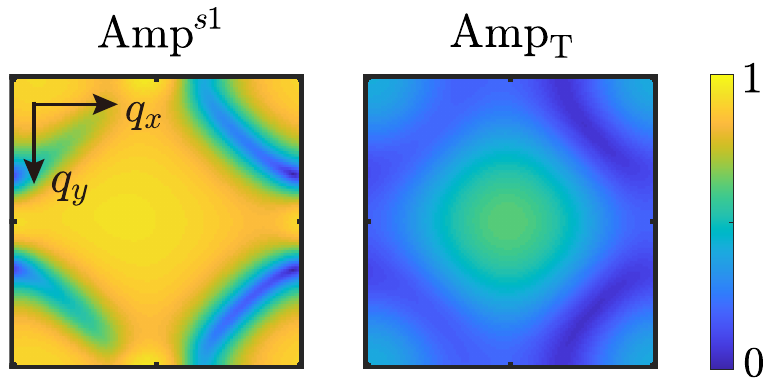}
 \caption{The amplitude of the oscillations for the $s1$ band $\text{Amp}^{s1}$ and finite temperature $\text{Amp}_{\rm T}$ in the FBZ.
 The asymmetry is from the coupling between the $s1$ band and other bands.
 Parameters: $(V_0,\Omega_0,\delta)=(4,1,-0.2)E_{\rm{r}}, t_{\rm{R}}=10\mu s,\delta_{\rm{R}}=0, \Omega_{\rm{R0}}\approx 3.4E_{\rm{r}}$.
}
  \label{FigAmpvsAmps1}
\end{figure}
Besides, we should also clarify that the higher-band population directly reduces the contrast of the spin textures and indirectly reduces the amplitude of the oscillations.
To this end, we generalize Eq.~(3) in the main text to a situation with finite temperature.
For finite temperature, the amplitude of the oscillations reads $\text{Amp}_{\rm T}=\sqrt{\langle \sigma_{z} \rangle_{\rm T}^2+(\langle \sigma_y \rangle_{\rm T} \cos\Delta\varphi+\langle \sigma_x \rangle_{\rm T}\sin \Delta\varphi)^2}$;
And the amplitude in $s1$ band $\text{Amp}^{s1}=\sqrt{(\langle \sigma_{z} \rangle^{s1})^2+(\langle \sigma_y \rangle^{s1} \cos\Delta\varphi+\langle \sigma_x \rangle^{s1}\sin \Delta\varphi)^2}$.
$\text{Amp}_{\rm T}<\text{Amp}^{s1}$ since $0<p^j<1$.
$\text{Amp}_{\rm T}$ and $\text{Amp}^{s1}$ with $\Delta\varphi=0$ are shown in Fig.~\ref{FigAmpvsAmps1}.
According to the calculation for the amplitude of the oscillations, a decrease behavior of the amplitude of the oscillations at finite temperature directly originates from the reduction of the contrast of spin polarizations.
However, the reduction of the contrast of spin polarizations is directly caused by the atoms populate in the higher bands.
In other words, the atoms populate in the higher bands indirectly reduce the amplitude of the oscillations.


\begin{thebibliography}{52}
\bibitem [{not()}]{note1}%
  \BibitemOpen
  \href@noop {} {}\bibinfo {note} {Compared with the definition of the quantum
  volume in Ref.~\cite{PhysRevB.104.045103,GoldmanSciPostPhys}, we have
  rescaled its coefficient (by a factor of $\pi$) here to facilitate a direct
  comparison of this quantity and the Chern number: this convention help to
  make the inequality look slightly more concise.}\BibitemShut {Stop}%
\end{thebibliography}
\end{document}